# Single-cell spatial (scs) omics
## Recent developments in data analysis

José Camacho[1,*], Michael Sorochan Armstrong[1], Luz García-Martínez[1], Caridad Díaz[2], Carolina Gómez-Llorente[3,4,5]

*Corresponding author: josecamacho@ugr.es

[1] Research Centre for Information and Communication Technologies (CITIC-UGR), University of Granada, Spain

[2] Fundación MEDINA Centro de Excelencia en Investigación de Medicamentos Innovadores en Andalucía, Granada, 18016 Spain

[3] Department of Biochemistry and Molecular Biology II, School of Pharmacy, Institute of Nutrition and Food Technology "José Mataix", Biomedical Research Center, University of Granada, Granada 18160, Spain

[4] Instituto de Investigación Biosanitaria, ibs.GRANADA, Granada, Spain

[5] CIBEROBN (Physiopathology of Obesity and Nutrition CB12/03/30038), Instituto de Salud Carlos III, Madrid 28029, Spain

# Abstract

Over the past few years, technological advances have allowed for measurement of omics data at the cell level, creating a new type of data generally referred to as single-cell (sc) omics. On the other hand, the so-called spatial omics are a family of techniques that generate biological information in a spatial domain, for instance, in the volume of a tissue. In this survey, we are mostly interested in the intersection between sc and spatial (scs) omics and in the challenges and opportunities that this new type of data pose for downstream data analysis methodologies. Our goal is to cover all major omics modalities, including transcriptomics, genomics, epigenomics, proteomics and metabolomics.

# 1 Introduction

Omics data refer to high-dimensional biochemical data at molecular level generated from highly sensitive and selective analytical instrumentation. Omics technologies are advancing rapidly, improving on throughput, reliability, and coverage in various domains. There are different technological trends to generate omics data based on the molecules of interest (i.e.: nucleic acids, proteins, metabolites, etc.). Methods based on sequencing [1] characterize nucleic acid sequences in the modalities of genomics, transcriptomics, and metagenomics. Spectrometric and spectroscopic technologies [2] enable characterization primarily in proteomics and metabolomics. Sequencing and spectrometry technologies have traditionally been developed independently, pushed forward by different technological communities.

Over the past few years, technological advances have allowed for measurement of omics features at the cell level, creating a new type of data generally referred to as single-cell (sc) omics [3], in contrast to the traditional 'bulk' omics that account for the total amount of omics features within a sample. sc omics technologies allow us to understand feature variations across cells, and to better infer cell-to-cell interactions. Alongside sc omics, the so-called spatial omics are a family of techniques that generate biological information in a spatial domain, of special interest to study the dynamics of biological conditions within tissues. Nature declared spatial transcriptomics method of the year in 2021 [4], and spatial multi-omics one of the seven technologies to watch in 2022 [5]. Spatial omics was also listed as one of the top 10 emerging technologies in 2023 by the World Economic Forum [6].

sc and spatial omics refer to two complementary characteristics: sc technologies do not necessarily provide information that is resolved in space [7] and some spatial technologies have resolutions lower than the cell level [8], [9], and so cannot be regarded as sc. Technologies which provide both omics features resolved in space and at sc resolution can be regarded as sc spatial (or scs) technologies. scs omics technologies undoubtedly represent one of the most powerful approaches today to understand disease propagation in both time and space [10], [11] while disentangling the complex biological interactions in differentiated cell populations [12], [13].

Omics data present several challenges for analysis and interpretation. Analytical instruments produce convoluted signals that require sophisticated pre-processing approaches, both for sequencing [14], [15] and spectrometry data [16], [17]. Furthermore, the biological variability of interest needs to be distinguished from confounding factors [18], [19] such as batch effects and other experimental factors and/or artifacts. These confounders can be more significant and numerous when combining multiple omics modalities [20]. Finally, capturing and integrating high resolution (e.g., sc) and/or spatio-temporal characteristics introduces additional challenges for the interpretation of experimental omics data [21], [22].

The evolution of sc and spatial omics has been heterogeneous across modalities, as illustrated in Figure 1. Research on spatially resolved features precedes that of sc resolution in most modalities. The fields of genomics and proteomics were the first to fully exploit spatial information (Figure 1(a)). Yet, it should be noted that the term 'spatial', when associated to different omics areas, has been used with disparate meanings, see for instance the discussion by Bouwman et al. about spatial genomics, 3D genomics and multiregion sequencing [23]. Figure 1(a) also shows that recent years have seen a surge of interest in spatial data from metabolomics but especially transcriptomics experiments. Regarding sc technologies (Figure 1(b)), all modalities have achieved an exponentially growing interest from the community, with sc genomics and transcriptomics now quickly becoming routine. If we combine both sc and spatial methods, the evolution in transcriptomics research is such that technologies in the field change so much from one year to the next, that there is a need for constant innovation in data analysis to derive the most accurate conclusions from the experimental evidence.

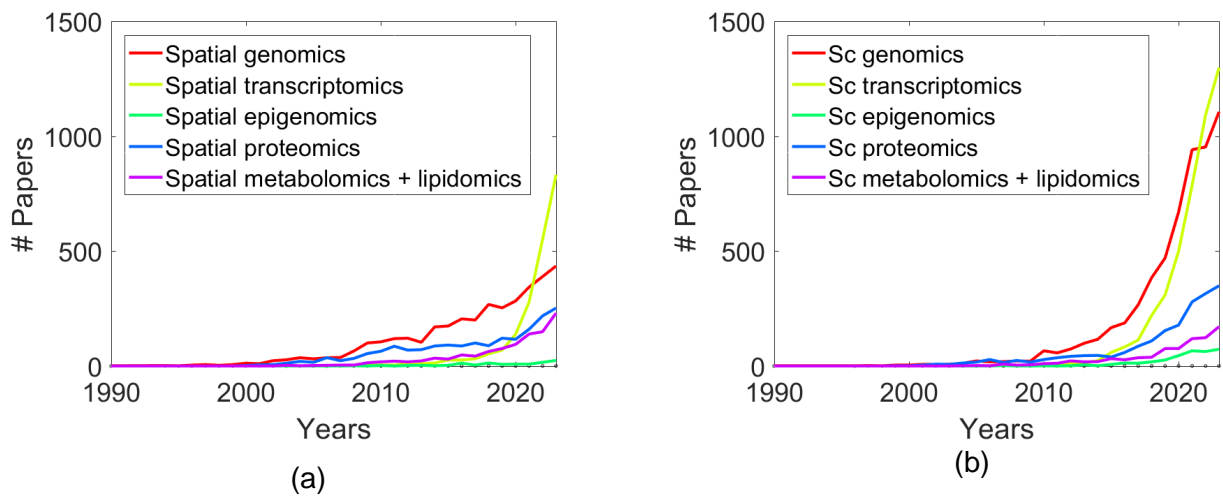

**Fig 1.** Number of papers published according to the omics modality: (a) spatial and (b) single-cell. Information extracted from the Web of Knowledge using the key words in the legend.

In this review, we are mostly interested in the intersection between sc and spatial (scs) omics and in the challenges and opportunities that this new type of data pose for downstream data analysis methodologies. Our goal is to cover all major omics modalities, including transcriptomics, genomics, epigenomics, proteomics and metabolomics. This review fills a gap of previous surveys, more concerned with sc technologies not necessarily spatially resolved, and in one or a subset of modalities [8], [24], [25], [26], [27]. The closest survey is by Vandereyken et al. [28], with an excellent discussion on the integration of different sc and spatial modalities. In comparison, our survey is more concerned with data analysis and machine learning tools, and with the challenges and opportunities to model the spatial information.

The rest of the paper is organized as follows. Section 2 contains a glossary of terms to clarify the discussion in the paper. Section 3 reviews the analytical technologies that produce scs omics in different modalities. Given the fast evolution in several modalities, especially spatial transcriptomics, we provide only a picture of current technologies and practices. Section 4 summarizes the major steps in the computational pipelines and discusses the structure of spatially resolved data. Section 5 describes common visualization techniques based on dimensionality reduction that are used for sc, spatial and scs data, with an emphasis on how spatial information is treated. Section 6 discusses approaches in Machine Learning that can be leveraged to model spatial data. Finally, Section 7 draws the conclusions.

## 2   Glossary of terms

**Omics features:** Relative quantification of biological units: genes, transcripts, proteins, metabolites, depending on the omics modality.

**Feature resolution:** The level at which different biological units may be properly discriminated against.

**Spatial resolution:** The finest spatial spot in which we can measure omics features (e.g., nm, µm, mm). The more spatial resolution, the smaller the spatial units for which omics measurements can be retrieved. Spatial resolution does not imply that the feature is spatially resolved (or spatially located) in a tissue. Spatial resolution is often specified in size levels, like at nm/µm/mm level, cell level, subcellular level, etc.

**Spatial localization:** The ability to locate a measurement. We can say that the measurement is spatially located or resolved. When applied to omics, it indicates that the measurement can be located in the surface or the volume of the biological sample. Spatial localization does not imply any spatial resolution.

**Cell size:** Cell size in the literature can often be understood as mass, volume, or diameter. In the context of sc omics, we are interested in the diameter, which is in equivalent units as the spatial resolution. Cell diameter depends on the cell type, ranging from 0.5 µm to hundreds of µm.

**Bulk omics:** The acquisition and analysis of omics features in the total sample. While this is the traditional approach to omics data, the name 'bulk' was coined recently and as the opposite to 'single-cell', implying the idea that bulk features are extracted as aggregated features in the entire population of cells.

**Single-cell (sc) omics:** The acquisition and analysis of the omics features per cell or, equivalently, at a spatial resolution equivalent to a cell size. Therefore, sc omics is concerned with resolution, but not localization.

**Spatial omics:** The acquisition and analysis of omics features resolved (located) in space. Therefore, spatial omics are concerned with localization, but not resolution.

**scs omics**: The acquisition and analysis of omics features at the cell-size resolution and resolved (located) in space.

**Modalities:** Alternative data structures collected from omics acquisition technologies, which can vary in the molecule of interest (genomics, epigenomics, transcriptomics, proteomics, metabolomics), in the data organization (bulk, sc, spatial, scs) and/or in the specific acquisition technology.

**Unimodal omics:** Omics data for a single modality, also referred to as single-omics or mono-omics.

**Multimodal omics:** Omics data for the combination of several modalities, also popularly referred to as multi-omics.

**Subcellular level**: Higher resolution than the cell, often to study the specific organization of molecules within the cell (e.g., in terms of the organelle where the feature is measured, or structural units in 3D genomics).

**Location reference system:** Location information associated to omics features can be provided in different ways: within one or several grid cells in a 2D/3D grid, within one or several pixels/voxels in an image, or within one biological unit (organelle, cell or region) in a biological network. The latter is the most useful but complex to gather, since it requires the localization of both the omics features during their acquisition, and of the biological units. The localization of features and biological units often requires the combination of different data sources.

**Spatially agnostic**: Computational algorithm/modelling method that does not consider the spatial localization of measurements.

**Spatially informed**: Computational algorithm/modelling method that considers the spatial localization of measurements.

# 3 Single-cell spatial omics technologies for data acquisition

New technical advances in DNA sequencing, oligonucleotide/probe synthesis, microscopy as well as spectrometric and spectroscopic techniques have enabled new scs technologies, capable of measuring the omics data at high resolution and resolved in space. We divide these techniques into two main groups: techniques for nucleotide detection (transcriptomics and genomics) and techniques for protein and metabolite determination.

## 3.1 Spatial transcriptomics and genomics techniques

As already discussed, most scs omics developments concerning the measurement of nucleic acids have been made on transcriptomics. Techniques vary in their spatial resolution, coverage, throughput and multiplexing capacity. Here we classify the scs omics technologies into two categories based on the main technique used to determine the transcript in the sample: microscopy or imaging methods and sequencing methods.

### 3.1.1. Imaging methods

These methods are based on Fluorescent *in situ* hybridization (FISH) and on single-molecule FISH (smFISH). These systems determine localization and relative quantification at the same time. Fluorescent probes hybridize to specific target nucleotide sequencing, enabling spatial localization. In addition, expression of a target nucleotide can be quantified from the fluorescence of the probe. Current FISH methods include sequential FISH (seqFISH), seqFISH+, multiplexed error-robust fluorescence *in situ* hybridization (MERFISH) and Enhanced

Electric Fluorescence *in situ* hybridization (EEL-FISH) [29], [30]. seqFISH+, MERFISH and EEL-FISH work at subcellular level and with a diffraction limited resolution.

seqFISH is a method in which fluorescent probes are designed to find and hybridize with complementary DNA (cDNA) or RNA sequences in the tissue sample. As there are only a few discernible fluorophore colours (four or five), applying sequential rounds of hybridization and imaging allows several hundred transcripts to be distinguished based on the sequential order of colours visualized by microscopy [31]. In seqFISH+ methodology, primary fluorescent probes contain four barcode sequences that serve as target sites for fluorescent secondary probes. Transcripts are identified in accordance with secondary probes bound to the barcode sites during four sequential rounds of probing, which is called barcoding [31]. Therefore, these modifications allow for the identification of thousands of molecules.

MERFISH is also built upon smFISH but it implements a combinatorial barcoding scheme where each target gene is assigned to a unique binary barcode. In general terms, thirty to fifty gene-specific barcodes probes hybridize to different positions on a targeted gene. Each probe also has hangout tails for readout binding. The fluorescent labeled or unlabeled secondary probes will recognize the tails of the first probes to generate the barcode as 1 (fluorescent) and 0 (not fluorescent). This method also employs several rounds of fluorescent hybridization and imaging [27]. MERSCOPE (Vizgen Inc,USA) was the first commercial platform based on this technology, launched in 2021.

EEL-FISH is an smFISH-based method that combines multiplexing with large-area imaging at high resolution. In this method, tissue RNAs are transferred onto a glass slide by electrophoresis and the tissue is removed prior to multiplexed FISH [29], [32].

Microscopy based methods can be used for spatial profiling of genome or epigenome, alone or in combination with transcriptomics by directly imaging DNA loci, chromosomal and nuclear structures. For example, MERFISH and seqFISH can be adapted to allow chromatic determination (DNA-MERFISH and DNA-seqFISH, respectively) [28].

### 3.1.2 Sequencing methods

In this category we can distinguish between Next Generation Sequencing (NGS) methods and *in situ* sequencing methods.

In NGS sequencing methods, the spatial localization of a specific gene or transcript is first recorded and subsequently the gen or transcript is sequenced. There are two main approaches: array-based approaches (10x Genomics Visium) and microfluidic barcoding strategies (DBiT-seq). These methods, in general, are not as efficient as FISH based methods [33].

In array-based approaches, first the tissue sections are stained and imaged using standard histological methods, and then the mRNA is captured on a microarray from fixed and permeabilized tissue sections. The microarray is coated with spatial barcoded RNA-binding probes. This spatial barcode acts as a messenger that is transferred to the cDNA during library construction, allowing localization of the transcript back to the tissue after sequencing. With this method, in transcriptomics we can only study poly(A)-mRNA. 10x Genomics Visium technology works at 50-μm resolution [27], [28].

Microfluidic barcoding strategies allow high-resolution profiling. DBIiT-seq (Deterministic barcoding) uses a microfluidic tool to create two sets of unidirectional massively parallel flows on a tissue piece. The first set provides DNA barcodes of primary (row) coordinates to the tissue-derived DNA, whereas the second sets are rotated (90º) from the first to provide DNA barcodes of the secondary coordinates (column). This yields a 2D map of pixels (10-µm pixel size) on the tissue with a unique combination to spatially allocate barcode NGS reads [28]. This technology can be used to locate transcripts, proteins or even (epi)genomic information. It is worthy to mention that DBiT-seq allows the integration of multi-omics spatial information [30].

In the field of epigenomics, some methods based on microfluidic barcoding strategies have been developed to determine chromatin accessibility or histone modifications, together with the transcriptome (ATAC&RNA-seq and spatial CUT&Tag-RNA, respectively) [34].

*In situ* sequencing methods enable spatial localization and sequencing at the same time. In these methods, tissues are fixed and RNA is converted into cDNA which are circularized and amplified by rolling circle amplification (RCA). The amplified cDNA are sequenced *in situ* using sequencing by ligation. In this approach, high-resolution imaging is required [30]. There are targeted and untargeted approaches for transcript identification, where specific probes hybridize to *in situ* generated cDNA, followed by probe ligation and RCA of either the barcode probe or short sequences of the cDNA, generating micrometre-sized RCA products inside the cell that are decoded using *in situ* imaging-based sequencing by ligation [28]. FISSEQ (Fluorescent *in situ* sequencing) and ExSeq (a combination of FISSEQ with expansion microscopy) use two bases per round. The detection efficiency of FISSEQ is lower than 0.005%. STARmap combines the *in situ* sequencing with hydrogel-tissue chemistry to develop a 3D intact-tissue RNA sequencing, where more than 1000 genes can be mapped [35]. These methods can also be adapted to (epi)genomic studies. Finally, oligoFISSEQ allows *in situ* sequencing-based visualization of multiple genomic loci [28].

Mathematical cartography is one of the last techniques developed in spatial omics studies. This method is based on imaging reconstruction by solving an inverse problem. DNA-GPS combines ideas from mathematical cartography and positional indexing. Briefly, after fixing cells, target RNA molecules are transformed into cDNA *in situ* by primer molecules encoding random short sequences as unique molecular identifiers (UMIs). The UMI-tagged cDNAs are amplified *in situ* and the amplified products are allowed to diffuse spatially. These cDNA amplicons are then fused with proximal amplicon products through overlap-extension resulting in two UMI-tagged gene products fused. The resulting random nucleotide pair at the junction is called a unique event identifier (UEI). The cDNA sequences, UMIs and UEIs are determined by DNA sequencing. This sequencing data allows the spatial positions of the RNA molecules to be reconstructed. DNA-GPS is a theoretical framework that combines mathematical cartography and positional indexing for large-scale optics free spatial genomics [30].

## 3.2 Spatial proteomics and metabolomics techniques

Advances in spatial proteomics and metabolomics or lipidomics have progressed steadily over the last 15 years. Spatial proteomics was introduced in 2009 in a study on *Leptospira interrogans* that combines quantitative mass spectrometric analysis using stable isotopes with image analysis by cryo-electron tomography with limited resolution [34]. Subsequently, spatial metabolomics began to develop. The techniques employed in spatial proteomics and metabolomics analysis differ primarily in sensitivity, specificity, resolution, and molecular coverage. Mass Spectrometry Imaging (MSI) is the technology of choice for scs metabolomics [8], [36] and proteomics [37], [38] as it best combines the previous characteristics. In the next subsections, we describe scs proteo/metabolomics methodologies divided into those using MSI and those based on other types of techniques.

### 3.2.1. Mass spectrometry-based imaging techniques

One of the most sensitive and selective categories of instrumentation in scs proteomics and metabolomics is MSI, which integrates the spatially resolved and high-resolution sampling of molecules from within a mass spectrometer. The collection is performed spatially by dividing the sample surface into a virtual grid of pixels. For every pixel in the grid, molecules are desorbed from the area of the pixel using a laser, ion beam or solvent, then a mass spectrum is generated to depict the relative intensities of the charged mass fragments within each pixel [39]. MSI imaging has demonstrated steady improvement in recent years in regards to spatial and mass spectrometric (feature) resolution, specificity in regards to interferents, and broader molecular coverage at the sample preparation stage.

Three major MSI methodologies are most commonly used: secondary ion mass spectrometry (SIMS), desorption electrospray spectrometry imaging (DESI), and matrix assisted laser desorption ionized (MALDI).

SIMS directs a high-energetic primary ion beam onto the surface of the sample, causing the ejection of atoms or molecules from the topmost layer. This step is followed by ionization/desorption and measurement to determine surface composition, chemical imaging and depth profile analysis, detecting the secondary ions generated through a mass spectrometer. SIMS has the highest spatial resolution of the three methodologies, with NanoSIMS reaching 50 nm (for elemental ions only), and provides further structural information by fragmentation [40]. In addition, chemical mapping of a 3D sample can be achieved by iteratively alternating between surface imaging and sputtering of the analyzed surface layer.

DESI generates charged droplets and ions of solvents through ESI, which are then propelled at a high velocity to ionize/desorb molecules in the gas phase. DESI offers analysis under ambient conditions from intact biological samples, without the need for complicated sample preparation. Nanospray DESI and laser ablation electrospray ionization (LAESI) are modifications of DESI which improve spatial resolution. Nanospray DESI employs a liquid micro-junction formed between two capillaries and the sample surface, where analytes are extracted and transported through the nanospray capillary into the mass spectrometer. LAESI is an ambient desorption/ionization technique ideal for samples containing water. It employs a mid-infrared laser with a wavelength of 2940 nm [41]. This laser excites water molecules, causing analytes to be ablated from the sample surface into the gas phase. The resulting gas plume then interacts with charged droplets generated by ESI, ionizing the ablated analytes.

MALDI is a technique with soft ionization that employs a laser beam and a photon-absorbing matrix to generate large molecular ions with relatively little fragmentation. To enhance MALDI, researchers have explored alternative methods, including matrix-free LDI techniques that employ mid-infrared lasers capable of vibrationally stimulating intrinsic water molecules within the cell. In this approach, probing water rich-samples allows the ablation of sample molecules, but sample post-ionization is necessary, since the water molecules themselves are mostly neutral. In infrared matrix-assisted laser desorption electrospray ionization (IR-MALDESI), an orthogonal electrospray is employed to intercept the ablation plume and direct the ions towards the mass analyzer [37]. Another approach to reduce ion suppression phenomena is based on a platform coupling visible-wavelength MALDI to an effective matrix (resorufin) with small background interferences. The visible-wavelength MALDI favors desorption and ionization, which allows a scs metabolomics and proteomics analysis with high sensitivity [42].

Mass cytometry is a variant of flow cytometry by mass spectrometry in which cells are isolated and target proteins are labeled with antibody markers, which in turn bind with various heavy metal ions of the lanthanide series. This methodology has recently been incorporated into high-dimensional imaging techniques using laser ablation, whereby subcellular resolution can be obtained [41]. The technique is aimed only at scs proteomics analysis.

The addition of ion mobility to mass spectrometry, which provides a collisional cross-section (CCS) bringing a fourth dimension in which ions are separated under a gaseous atmosphere through the influence of an electric field, is critical to enhance the accuracy of protein, peptide, metabolite and lipid assignments and improve the quantification completeness and metabolite coverage for scs analysis.

Previous methodologies locate the omics features into the image grid. Some recent technological developments combine this localization with cell segmentation: the Space M platform uses light microscopy for image segmentation and calibration, followed by MALDI metabolomics imaging. With this approach, scs analysis can be performed with high-throughput, obtaining information for more than one hundred metabolites [43]. Recently, the combination of untargeted spatial metabolomics and targeted multiplexed protein imaging in a single pipeline has been possible with Single Cell Spatially Metabolic (scSpaMet), fusing 3D spatially resolved metabolomic profiles acquired by time of flight (TOF) SIMS and mass cytometry proteomics images. scSpaMet introduces analysis capabilities for joint metabolomic and proteomic single-cell data [42].

### 3.2.2. Techniques not based on mass spectrometry

There are several non-MSI techniques to explore metabolites or proteins at scs level. Fluorescence lifetime imaging (FLIM) is a non-invasive modality that measures the fluorescence decay of endogenous fluorophores [44]. Another technique is Fourier-transform infrared microspectroscopy (FTIR), with the ability to differentiate biochemical changes at the proteins and lipids level [45]. In addition, the combination of Raman spectroscopy with microscopy can be used for scs analysis [12]. Other approaches used in proteomics are based on specific antibodies that bind to proteins and make use of immunocytochemistry and immunohistochemistry techniques that measure fluorescence, or DNA antibody tags, based on sequencing methods involving antibodies fused to DNA oligonucleotides [28]. This type of analysis has limitations due to lack of specificity.

# 4 scs omics preprocessing

Given that technological trends in bulk, sc, and scs omics vary depending on the molecule of interest (nucleotide, protein, or metabolite), computational pipelines need to be adjusted for these differences. Several recent surveys discuss the main steps in the computational pipeline of scs omics for a single modality [9], [28], [46], [47], [48]. Those surveys provide a detailed discussion about technological and data analysis differences across methods, as well as available software tools that the analyst may resort to. Here, we provide a global view to pipelines of different omics modalities, trying to spot similarities and differences that may help foster the cross-fertilization of ideas and processing techniques. The remainder of this section is concerned with the necessary steps to retrieve the scs information and the resulting data organization. Approaches to visualize these data and use them in further applications are treated in next sections.

## 4.1 Computational Pipelines

As we have already mentioned, technological advances in the determination of scs nucleotide material are dominated by the field of scs transcriptomics. The computational pipeline in scs transcriptomics is very dependent on how the spatial location is retrieved from a sample. To that regard, there is a fundamental difference between *in situ* methods (both *in situ* hybridization and sequencing) and NGS methods [9]. *In situ* methods determine the spatial location as a pixel in a microscopic image, while in NGS methods the location is determined in a 2D grid using barcode sequencing. In both cases, the goal is to identify the transcript locations within the cell, and the cell within the tissue under study.

The computational pipeline for *in situ* methods start with a preprocessing step in which raw images with the location of transcripts are aligned and the signal-to-noise is enhanced. Then, transcripts are identified and quantified by pixel/voxel [9]. The result is a tensor representation of the data with omics-per-pixel/voxel, as described later in Section 4.2. This tensor has one mode for the omics features (the transcripts) and as many modes as dimensions in an image. Yet, this tensor is not useful in several applications, like to infer cell-type distributions and cell-cell interactions. By performing spatial segmentation, we assign the transcripts to biological units (e.g., cells). This segmentation can be done using image processing techniques [49] or spatial distribution models for the transcripts [50]. The result is a couple of data structures that complement the previous tensor, with information about the cell locations and with counts of omics-per-cell.

The computational pipeline for NGS methods starts in a similar way as regular scRNA-seq. Locations are sequenced together with transcripts, which are quantified and stored in an array of locations (omics-per-grid cell). Often, the resolution of the array is limited, and deconvolution techniques are applied to enhance it. Finally, spatial segmentation is performed in a similar way as before, to assign transcripts to biological units (omics-per-cell).

Popular tools for scs transcriptomics are Seurat [3], [51], an R package designed for quality control, analysis, and exploration of NGS sc/scs transcriptomic data and for combination with other data modalities, Squidpy [52], within the Scanpy [53] python package, and Giotto python package [54]. A complete list of tools is provided elsewhere [9], [46].

MSI for scs metabolomics and proteomics bears more similarity with *in situ* transcriptomics methods. The first step is image preprocessing for signal-to-noise enhancement [48], including baseline correction, noise-reduction filtering, spectral alignment and normalization. In a similar fashion as for transcripts identification, the next step is peak selection, often performed taking advantage of the shape of peaks. Binning can be performed either before or after peak selection. Another relevant step is protein/metabolite annotation and identification [49]. In a substantial difference to spatial transcriptomics, MSI spatial data is often analyzed in the original image coordinates (omics-per-pixel), rather than compartmentalized in biological units. This is, nonetheless, problematic for the integration with other scs modalities [28]. Recent research looks at cell segmentation and association to omics features (omics-per-cell) using microscopic images, performed similarly as in scs transcriptomics [43], [55]. A complete list of tools for scs metabolomics and proteomics [48], [49], and multi-omics [28] can be found elsewhere.

## 4.2 Spatially resolved data

In the remainder of the paper, we identify scalar elements with lowercase letters, array dimensions with uppercase letters, vectors (1D arrays) with bold lowercase letters, matrices (2D arrays) with bold uppercase letters, and tensors (high-order arrays) with underscored bold uppercase letters.

Traditional (bulk) omics studies are performed on a sample-wise basis, so that a single vector of omics features is generated per biological sample. Data corresponding to a complete study can be represented with a data matrix **X** with I rows and J columns, with I the number of experimental units (samples, individuals) and J the number of omics features[1] (Fig 2.A). For multi-omics experiments with the same individuals, data may be stored in several matrices for each omics modality, with I rows and different columns: $J_1$, $J_2$, …, $J_m$ (Fig 2.B). sc omics can be represented in a similar matricial form as bulk omics, but now the rows correspond to the cells per sample. This means we obtain a single matrix (for a single modality) or several matrices (for multi-omics) per biological sample, rather than a vector.

---

[1] For the sake of homogeneity, we assume that experimental units are stored in the rows and omics features in the columns. Note however that some researchers may be more familiar with the inverse organization.

In the case of spatial omics, in its original form (omics-per-grid call, omics-per-pixel/voxel), one sample corresponds to several feature vectors across 2 or 3 spatial dimensions. This provides richer information about spatial heterogeneity, but introduces new problems related to coverage and data labelling, among others. Spatial omics data can be stored in a tensor $\underline{\mathbf{X}}_s$ with I individuals, J omics features and S locations in a 2D (or 3D) grid with the spatial dimensions: $S_1$, $S_2$ (and $S_3$). For instance, in the example of Fig 2.C, $\underline{\mathbf{X}}_s$ is a 4-way tensor of dimension I x J x $S_1$ x $S_2$, corresponding to spatial omics in a 2D grid. Notice this structure assumes that the spatial distribution across the I individuals is comparable, that is, that the corresponding scs omics samples can be transformed to a common coordinate system, which is a strong requirement. The recent development of alignment algorithms for this purpose [56], [57], [58], [59] represents a cornerstone to make this configuration useful to investigate spatio-temporal patterns associated with disease.

Reflecting on the final tensor structure in $\underline{\mathbf{X}}_s$ allows us to better understand how data models capture the spatial information. This tensor representation is suitable for spatial omics data regardless of the resolution.

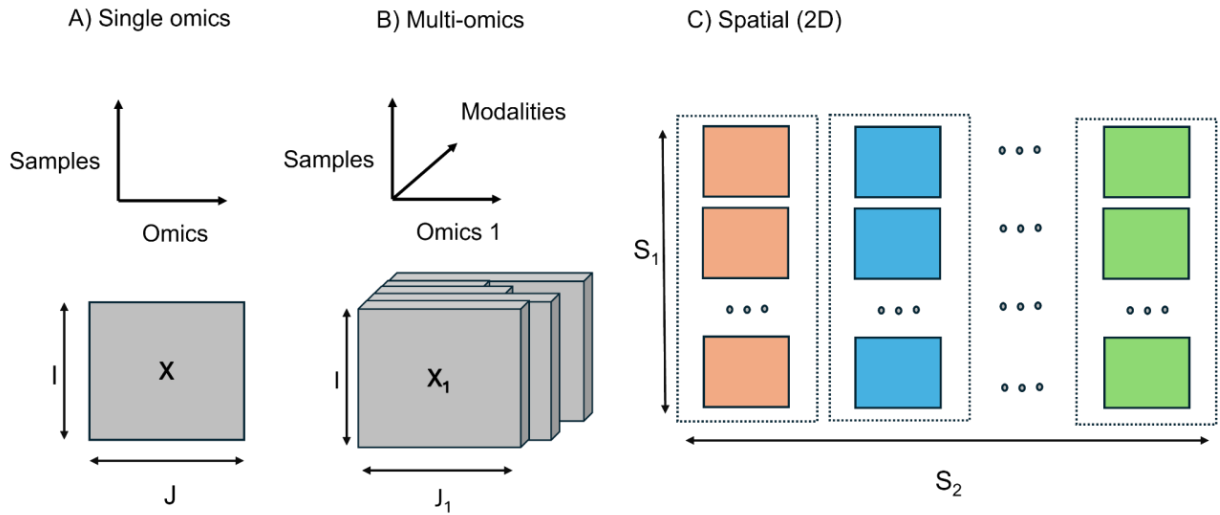

**Fig 2.** Data structure for A) omics with one modality, B) multi-omics, (C) spatial omics in a 2D grid

The omics-per-cell data can be represented in a matrix with I cells and J features, just like in sc data. Yet, this matrix does not provide spatial information. Given that the cells have uneven volumes, the omics-per-cell data cannot be represented directly as a tensor like in $\underline{\mathbf{X}}_s$. There are two simple solutions to handle the spatial information in the omics-per-cell data [54], to transform the data back to a coarse grid, or to build a neighbourhood network, where each cell is represented as a node, and neighbouring cells are connected with an edge. Methods based on network graph theory [60] are extensively used in molecular biology, with biological units typically described as *nodes* or *vertices* in a graph, and connections between these nodes calculated through some distance metric.

# 5 Data visualization and analysis by dimensionality reduction

Visualization of omics data, necessary for data interpretation and hypothesis generation, is central in downstream data analysis. Dimensionality reduction methods map the data onto a lower-dimensional (latent) space (usually 2D, sometimes 3D) for visual representation, since the omics features may number in the thousands and more. Methods vary with respect to the optimization criteria, assumptions, flexibility and interpretational capabilities, as illustrated in Table 1. In the context of sc and scs omics data, the most popular visualization approach is the Uniform Manifold Approximation and Projection (UMAP) [61]. Yet, we can generally say that there is no single method that suits all datasets and visualization / investigation needs, so it is interesting to reflect on the advantages and disadvantages of the most extended visualization approaches.

We would like to make a distinction between spatially agnostic methods, those methods that do not use location information into the dimensionality reduction process, and spatially informed methods which encode location information as part of the cost function of the algorithm for dimensionality reduction. Spatially agnostic methods have been traditionally used for any type of multidimensional data, including bulk and sc omics data. Spatially agnostic methods can be used to visualise scs data in two steps: i) dimensionality reduction is computed in a manner which is agnostic to the spatial information, and ii) the result is projected in a tissue image to create a visual map. Most visualization methods in the spatial/scs omics literature are indeed spatially agnostic, which has the following (often unperceived) benefit: all spatial patterns that we find in a visual map made using a spatially agnostic method can be essentially trusted, given that the method was not actively looking for them (that is, we are not pushing the projection to be smooth in space). Contrarily, all spatial patterns in a visual map made with a spatially aware method need more careful validation. This is akin to what happens with unsupervised and supervised methods in machine learning. There is an overwhelming usage of spatially agnostic methods in sc and scs.

If the analysis involves multiple modalities, either each modality is treated on a separated basis or the data from different modalities is combined. If the individuals in each modality are consistent, omics features from different modalities can be appended along the omics mode (Fig 3.A). Yet, most of the time the combination is not straightforward, and we can resort to data fusion techniques [62] or enrichment analysis [54].

Regarding the spatial information, spatially agnostic processing techniques can be applied over the tensor $\underline{\mathbf{X}}_s$ (omics-per-pixel/voxel/grid cell) by conveniently unfolding this tensor to 2D first. This is the standard spatially agnostic approach for MSI data: each pixel/voxel/grid cell is unfolded along a single sample mode (from Fig. 3.B to Fig. 3.C). Combining the two unfolding operations in Fig. 3 result in a matrix $\mathbf{X}_s$ with $I_s = I \times S_1 \times S_2 (\times S_3)$ rows and $J_s = J_1 \times J_2 \times \ldots \times J_m$ columns. In scs transcriptomics and related analyses, the matrix of omics-per-cell (with I cells and $J_s$ columns) is most often used for visualization.

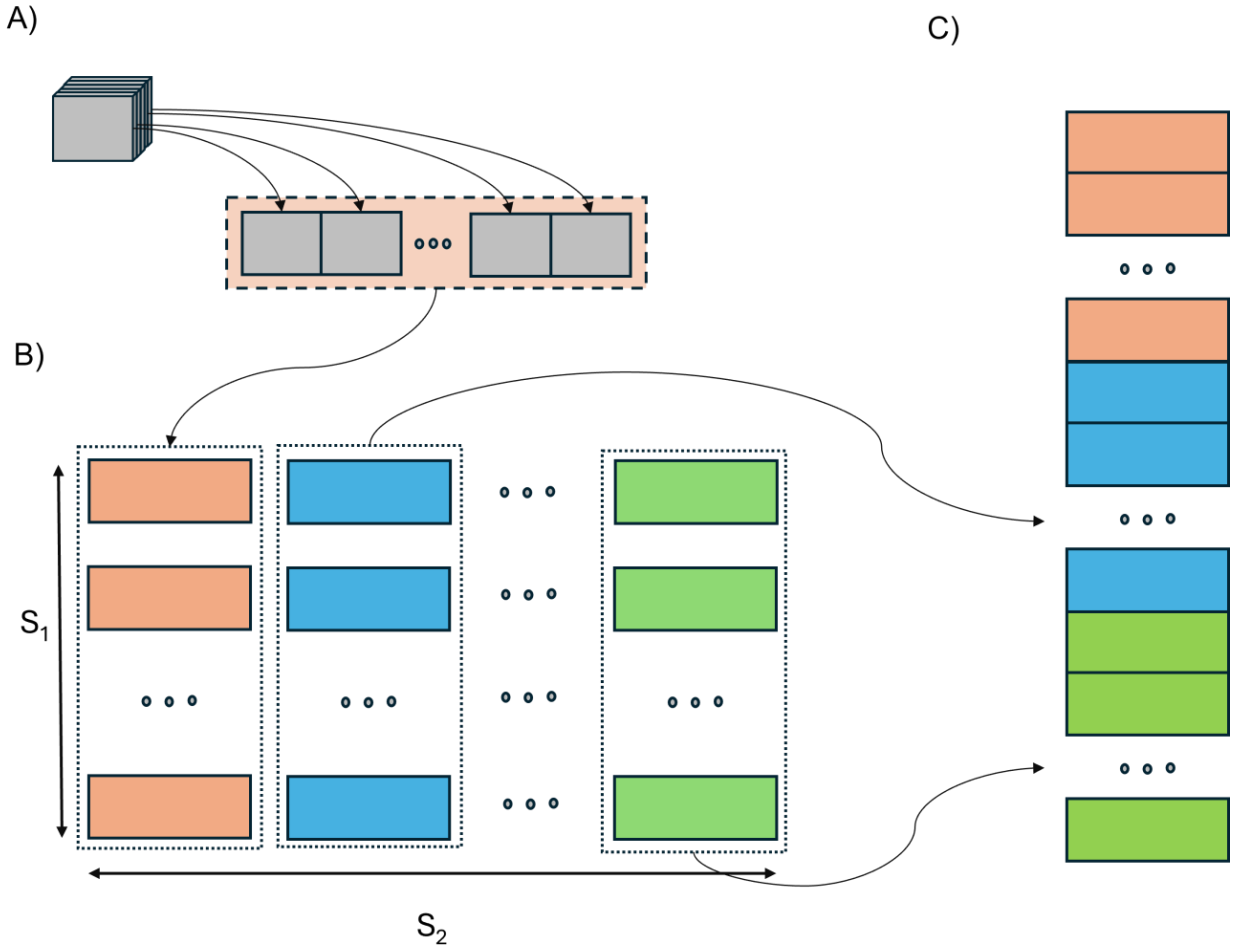

**Fig 3.** Data structure for A) multi-omics data unfolded into a single omics mode, (B) spatial multi-omics in a 2D grid and (C) unfolded multi-omics data in a single sample mode.

## 5.1 Spatially Agnostic Methods

Let us consider a matrix $\mathbf{X}$ with I rows and J columns that represents a number of I experimental units (samples, individuals, cells) where J omics features are stored (from one or several modalities). Since we are discussing spatially agnostic methods, we make no distinctions if $\mathbf{X}$ comes from traditional (bulk or sc) omics experiments or from spatially resolved omics (so that $\mathbf{X} = \mathbf{X}_s$ after the unfolding operation in Fig. 3, or represents the omics-per-cell matrix). Sometimes we can also consider another block of data $\mathbf{Y}$ with I rows and K columns, which can represent other sources of data for the same objects in the rows (e.g., the anchors in Seurat [3] to combine multiple sc data sets), or a set of categories/classes associated with the rows (e.g., case vs control). Note that general procedures for data fusion within the dimensionality reduction framework are treated elsewhere [62].

| Method | Input | Meta-parameters[2] | Optimization | Useful to Visualize | Assumptions | Flexibility | Interpret Rows in X | Interpret Columns in X | References for tools |
|---|---|---|---|---|---|---|---|---|---|
| **PCA** | Data matrix **X** | A | Variance (SS) within **X** | Patterns of high variance in **X** | Linear subspace | Middle | Yes | Yes | [51] [52] [53] [54] |
| **CCA** | Data matrices **X** & **Y** | A | Correlation **X** & **Y** | Patterns in **X** and **Y** | Linear subspace | Middle | Yes | Yes | [51] |
| **PLS** | Data matrices **X** & **Y** | A | Covariance **X** & **Y** | Patterns in **X** predictive to **Y** | Linear subspace | Middle | Yes | Yes | [47] |
| **PLS-DA** | Data matrix **X** & row classes | A | Covariance **X** & **Y** for **Y** dummy | Patterns in **X** predictive to **Y** | Linear subspace | Middle | Yes | Yes | [47] |
| **Sparse Methods** | Data matrix **X** or data matrices **X** & **Y** | A, #NZE | Toff Loss of Method + Sparsity | Patterns of high variance in **X** with low NZEs | Linear and sparse subspace | Low (const.) | Yes | Yes | [51] |
| **NNMF** | Data matrix **X** | A | Loss of Method for only Positive | Non-negative patterns of high variance in **X** | Linear and non-negative subspace | Low (const.) | Yes | Yes | [47] [63] [64] |
| **MDS** | Distance matrix from **X** | DT, A | Global (Dis)Similarities in X | Global patterns of (Dis)similarities | Global distance | Mid. / High | Yes | No* | [65] |
| **t-SNE** | Distance | DT, | Toff Global + | Complex | Symmetri | High | Yes | No* | [51] |

---

[2] Preprocessing approaches prior to model computation are not considered as metaparameters.

|  | matrix from **X** | A, P, E | Local (Dis)Similarities in X | patterns of (Dis)similarities | cal relationships between observations |  |  |  | [52] [53] [54] |
|---|---|---|---|---|---|---|---|---|---|
| **UMAP** | Distance matrix from **X** | DT, A, NN, MD | Toff Global + Local (Dis)Similarities in X | Complex patterns of (Dis)similarities | Uniform distribution on manifold | High | Yes | No* | [51] [52] [53] [54] |

**Table 1:** Dimensionality reduction methods for visualization. PCA stands for Principal Component Analysis, CCA for Canonical Correlation Analysis, PLS(-DA) for Partial Least Squares (Discriminant Analysis), NNMF for Non-Negative Matrix Factorization, MDS for Multidimensional Scaling, t-SNE for t-distributed Stochastic Neighbor Embedding and UMAP for Uniform Manifold Approximation and Projection. A: dimensionality of the projection subspace (number of latent variables). #NZE: number of non-zero elements in model parameters estimates. DT: distance type. Toff: trade-off. No*: information about columns and or between rows and columns of X may still be estimated with approximate techniques. P and E refer to metaparameters that are specific to t-SNE: Perplexity (related to the number of nearest neighbours in the graph representation) and Exaggeration (related to tendency of the data to cluster in the lower-dimensional representation). NN and MD are the number of nearest neighbours that span the local graph representation, and the minimum distance of the observations in the low dimensional representation for UMAP.

There is a large number of dimensionality reduction visualization techniques. Some of the most widely used in omics data are described below and summarized in Table 1. Several methods are based on the direct analysis of **X** (and **Y** if the approach considers more than one block of data). This includes Principal Component Analysis (PCA) [66], Canonical Correlation Analysis (CCA) [67], Partial Least Squares regression (PLS) [68] and Discriminant Analysis (PLS-DA) [69], and constrained versions of the former, such as sparse methods [70], [71], [72] and Non-Negative Matrix Factorization (NNMF) [73]. Very popular methods, though, work on a distance matrix computed from **X**. Distance is most often computed along the feature mode, so that the matrix contains information about the (dis)similarities among the objects (rows) in **X**, which is an effective approach to find clusters of objects and a popular approach to find different types of cells in sc omics. Methods of this sort include Multidimensional Scaling (MDS) [74], t-distributed Stochastic Neighbourhood Embedding (t-SNE) [75] and UMAP.

PCA is a standard tool that makes no assumption whatsoever on the data characteristics. The projection subspace is chosen to maximize the variance explained by the model, which allows us to visualize large patterns of heterogeneity across rows and columns in the data. For these two previous reasons, no assumptions made and a focus on high variance, PCA represents the multivariate workhorse for visualization: a general approach to let the data speak that represents a suggested first visualization step. Its main limitation though is that it projects the data in a linear subspace. Therefore, while no assumptions are made on the data themselves, the assumption of linearity is indeed made when calculating the projection subspace. Linearity has the nice property of preserving the original distances within the projection subspace, but this feature is often perceived as a lack of flexibility, since relevant patterns (like those that differentiate some cells from others) may not be obvious within a linear subspace that is optimized with respect to variance.

To find more subtle patterns, we can either avoid the linearity constraint and/or relax the focus on high variance within the **X** block. The latter can be done by using a supervised approach, like in CCA and PLS(-DA), which incorporate another block (**Y**) in the analysis. This second block can be used to provide information of the patterns we are looking for, or to combine the information gathered from two different omics modalities. CCA and PLS(-DA) still project the data in a linear subspace, but they can find useful information using a much lower fraction of variance. Yet, as supervised approaches, these methods are prone to overfitting (they can find seemingly interesting patterns from even random data), which means that careful validation must be performed [76].

Some constrained versions of PCA and other multivariate methods, like sparse methods and NNMF, represent an effective way of finding patterns with specific features. Sparse methods combine variance maximization and sparsity with the ultimate goal of improving data interpretation. In this context, sparsity is imposed by taking as many model parameters to zero as possible, so that interpretation is focused on the remaining. Non-negative factorizations constrain model parameters to be non-negative (either zero or positive), also to improve interpretability. These methods are quite inflexible in the patterns they can find (the more constraints - the less flexibility), so one needs to have a good intuition that the applied constraints are consistent with the data or at least suitable for the analysis goals. Sparse methods are especially interesting for some forms of sc/scs data [77]. On the other hand, omics features are universally positive, and non-negativity constraints may better facilitate a high-fidelity interpretation.

All previously described projection approaches, at least in their most rudimentary form, assume a linear projection subspace. It is worth commenting that this characteristic has the (often ignored) advantage of providing full interpretation of the feature mode in connection with the observation mode, so that we can identify which omics features are related to differences observed across cells. That may be especially relevant in the context of biomarker discovery [47].

A second type of visualization tools focus on the distance between pairs of rows of **X**, according to a specific distance criterion. This has two consequences. The first (and best known) consequence is that the introduction of the distance allows for further flexibility, including the modelling of non-linearities, since a wide range of distances (Euclidean, Mahalanobis, Hamming, Jaccard, etc.) can be used, changing the result of the visualization. This can also introduce a complication, since an additional choice, the distance, must be done in the data pipeline. The second consequence is that when we compute distances across the sample mode of **X**, the link between the distance and the corresponding feature mode is lost in the process, and therefore the visualizations only provide useful information about the distribution of the rows in **X** but lose how these directly relate to the original omics features (which can only be estimated). Again, this makes biomarker discovery more difficult. There is a variety of MDS algorithms for different data and distance characteristics, including Principal Coordinates Analysis (PCoA) [78], which is equivalent to PCA when the euclidean distance is used, and metric and non-metric MDS. The flexibility of manifold learning techniques such as MDS can be further capitalized upon in methods like t-SNE and UMAP, that optimize a subspace with respect to local distances between observations to find hidden patterns which are not apparent when only global distances are considered. t-SNE has long been employed in scRNA-seq experiments, because of its ability to capture local structure in the data in a way that best summarizes some of the most interesting trends in a lower-dimensional visualisation [79]. It has also been generalized to multiple modalities [80]. UMAP has been widely utilized in sc/scs experiments owing to its computational performance relative to t-SNE, and greater flexibility for handling some finer structures in the data. It has also been generalized to the multimodal case. UMAP has been used extensively as an untargeted method for the analysis of spatial omics data. Smets et al. applied this approach to MSI data [81] on pancreatic tissue samples, and compared the results with PCA and t-SNE using a variety of different metrics. UMAP was shown to be more computationally efficient than t-SNE, and competitive in terms of what latent structures were resolved. The study highlighted that the use of distance metric greatly impacted the findings of the study.

As expected, the relative usage of the data visualization tools is different in the areas of scs genomics and transcriptomics than in scs proteomics and metabolomics. In transcriptomics, Seurat includes PCA, t-SNE, UMAP as main visualization tools, and CCA for integration with a reference. In the analysis of MSI data, PCA and PLS-DA are the most popular methods, but other methods are also used [81], [82], [83].

## 5.2 Spatially Informed Methods

There has been a recent interest on the derivation of spatially informed methods that make the most of the distribution of spatial and scs omics for dimensionality reduction. SpatialPCA is derived from a probabilistic model for PCA based on an *a-priori* distribution of residuals [84]. SpatialPCA employs Probabilistic PCA with a Gaussian kernel matrix to explicitly model the spatial correlation structure across tissue locations. The intuition behind that choice is that neighboring locations are expected to be similar to each other. Nonnegative Spatial Factorization (NSF) extends NNMF with the spatial information in a similar way [85]. SpiceMix is an algorithm that incorporates spatial information via a hidden Markov random field (HMRF) graph representation, where neighbors are defined from an undirected graph computed from the locations of the spatial observations. The HMRF graph informs an NNMF representation of the data [86]. The Voyager scs package [87] includes MULTISPATI PCA, a combination of PCA with Moran's I, a spatial autocorrelation metric which grows with the correlation of the signal among nearby locations in space.

While the number of spatially informed approaches in scs is rather limited, there are related areas of research with a focus on the analysis and visualization of (often highly) multivariate data that is spatially resolved. This is the case of Multivariate Image Analysis (MIA), Multivariate

Curve Resolution (MCR) and spectral analysis. We can find inspiration in those areas to develop new computational methods for scs omics data.

MIA refers to the use of multivariate analysis in multi-channel images, from RGB (or other color code) to hyperspectral images [88]. In MIA there is always a tradeoff between the modelling of spatial information, through the so-called texture analysis, and channel information. MIA methods can work at pixel level (or voxel level for 3D) or at global level. The basic difference is how data is rearranged from the 2D/3D grid structure to a matrix/tensor for the subsequent analysis and visualization, in a very similar fashion as what is depicted in Figs. 2 and 3. Prats-Moltanbán et al. [88] discuss various approaches that go from a spatially agnostic approach equivalent to Fig. 3.C, to global approaches where the spatial information is completely encoded within the feature mode using a combination of PCA and the Discrete Wavelet Transform (DWT). An intermediate case is to embed the traditional idea of local windows in image processing into the unfolding/rearrangement process [89].

A major application in MIA is the analysis and visualization of hyperspectral images (HSI), where the different channels correspond to different wavelengths across the electromagnetic spectrum. HSI literature is concerned with new methods to combine spatial and spectral information. When the hyperspectral image is derived from a mixture of chemical components, the final goal is often to derive both the (pure) spectra of the components along with their spatial distribution in the image using MCR techniques [90]. A cornerstone of MCR methods is the definition of modelling constraints that encode chemical and physical assumptions [91]. MCR can be readily applied to MSI data from spatial proteomics and metabolomics in a spatially agnostic way [92], similar to that discussed in previous sections. Yet, there has been a recent interest in leveraging the spatial information in the form of additional constraints, achieved through data reorganization [93] or explicit definitions [94], [95].

Frequency spectrum analyses encompass techniques that encode spatial information as an abstract, $N$-dimensional representation of oscillating components with known periodicity. This is calculated over the entire signal, in the case of Discrete Fourier Transforms (DFTs), but may also be done incorporating an additional spatial component in the case of the Discrete Wavelet Transform (DWT). Both techniques have been widely deployed for MIA [96].

The most typical methods for spectral analysis are the Fast Fourier Transform (FFT) and inverse Fast Fourier Transform (iFFT), which are appropriate for experiments with regular sampling in both spatial coordinates (like in the omics-per-pixel/voxel/grid cell tensor). The power spectrum is calculated from the squared Fourier coefficients, incorporating the real and imaginary components of the series of complex numbers as an estimate for the relative energy expressed at each frequency. The power spectrum has been analyzed directly for single cell omics studies [97] as a feature extraction step, but has been additionally encapsulated within a tensor learning framework to represent spatial information within the context of high-dimensional omics data [98]. Although relatively limited in the omics information provided, a similar type of analysis has been successfully deployed to compare confocal microscopic images of amyloid plaques in post-mortem brain tissue based on an untargeted analysis of the fluorescence spectra, as something reminiscent of a spatial proteomics analysis [99].

Spatial information in the omics-per-cell structure (e.g., the neighbouring network) may be represented as irregularly sampled arrays through a Non-Uniform Fast Fourier Transform [100], but the inversion of such a transformation is challenging [101] and relatively few software libraries exist to support these types of operations with modern numerical software.

## 5.3 A note on experimental design

To answer a biological question with scs experiments, the experimenter needs to make a number of choices, including the type and size of tissue, the technology or technologies to use, which introduce some requirements about the preservation and preparation of the sample, the number of samples and the number of omics features, and whether the analysis is targeted or not. Providing best laboratory practices is out of the scope of this paper and can be found elsewhere [27]. Our interest here is to reflect on how the experimental design and the analysis depend on each other.

From the seminal work of Fisher at the beginning of the XX century, it is well-known that complex experiments need to be carefully designed to maximize the statistical power of the analysis approach. Statistical experimental design adds another dimension to laboratory considerations on the design of experiments, introducing effective design tools (like randomization, replication, and blocking) [102] to minimize the influence of confounders on the analysis results [18], [19]. This is especially necessary for scs experiments, given the sparse and extremely massive nature of resulting data, which makes statistical inference a real challenge [21], [22].

A powerful approach to data analysis and statistical inference of complex omics data is to explicitly integrate information about the experimental design (relevant covariates, crossed and nested factors, random factors) into the data analysis tool. It is widely accepted that this integration improves the statistical power. Powerful analysis tools that integrate this information include PERmutational ANOVA (PERMANOVA) [103] and ANOVA Simultaneous Component Analysis (ASCA) [104]. Furthermore, recent power analysis approaches specifically devised for these methods have been defined [105], [106], with which the number of samples required for a given statistical power can be estimated a-priori. Yet, none of these tools have been extended to handle the sparse, massive and spatial structure in scs data.

# 6 Modeling the spatial information with Machine Learning

## 6.1 Applications of machine learning to scs omics

Algorithms for the analysis of scs omics data may leverage what can be generally categorized as Machine Learning (ML) at various stages of the analysis pipeline. Associating multiple subcellular measurements into individually segmented cells for instance may utilize deep learning methods [107]. Owing to the flexibility through which multiple modalities of data can be incorporated into a single model, such a model can be used to perform multiple functions that incorporate information from several different biological domains [108]. In instances where these biological domains are collected on separate instruments, or experiments, a notable application of ML is to align the multi-modal data along a consistent spatial modality [109].

A classical feature of ML based methods is clustering, whereby different tissues may be identified as discrete characteristics of the data worth examining and interpreting separately. These clustering methods may be performed agnostic to spatial information or incorporating spatial information - and this may be done before or after dimensionality reduction to improve the computational efficiency of the routine in the case of high-dimensional data. In the case of deep learning, the activation of any neuronal layer can be visualized on a per-sample basis using UMAP to examine the differentiation of individual cells as a function of the learning mechanism/activation [110]. This is often used to describe the "latent space" of the neural network, and can be used to evaluate its ability to differentiate between different tissues according to a known criteria such as histological information. Clustering information, regardless of the source, can be used to inform spatially variant genes (SVGs) in the case of spatial transcriptomics and genomics, through a subsequent analysis of differential expression (DE) [111]. This analysis can be used to ascribe certain clusters to known cell types from previous experiments, or to infer mechanisms for Cell to Cell Communication (CCC) [112].

At the intersection of sc omics and scs omics is the ability of ML models to incorporate spatially agnostic data at the cellular level, into spatial data at a lower resolution to either up-sample the spatial information (resolution enhancement) [113] or estimate the relative population of cells at the spot level (deconvolution) [114].

## 6.2 Machine Learning strategies for scs

There are several interesting surveys on the intersections between spatial ML and ssc omics [22], [115], [116]. This section is based on the idea that incorporating spatial information into the ML models improves their performance and usability in terms of biological interpretation.

Let us consider spatial omics data as a stochastic process defined by a collection of random variables (omics features) indexed by points in space. There are 3 distinctive structural properties that differentiate spatial processes from other types of stochastic processes [117]. Firstly, the property of spatial dependence ("close entities are more related than distant entities") applies, violating the assumption of i.i.d. observations. This consideration might enhance the learning process. Spatial weights matrix, spatial kernels, spatial association rules or graph-based models (see section 6.2.1 and 6.2.2) are examples of using the spatial dependency property to model scs data. Spatial dependency is an intrinsic property of scs applications.

Secondly, a spatial stochastic process likely presents a certain degree of spatial heterogeneity defined as being spatially non-stationary. That is, the difference between random values (in omics features) in different spatial points will not only be a question of spatial distance, since spatial dependencies are anisotropic (non-uniform in different directions) [117]. Addressing the spatial heterogeneity in spatial omics data is a powerful tool for, e.g., quantitative description of microenvironments defined as vectors of cell type frequencies within cellular neighborhood, spatial domain identification, or SVG detection among many others. Spatial dependence and heterogeneity are properties that can only be leveraged by spatially resolved omics techniques (spatial and scs technologies). To that regard, scs ML approaches can also be split into spatially aware approaches and spatially agnostic approaches.

The third structural property of spatial data is the scale. The scale of the data stochastic process is concerned with the resolution (measurement scale), context (scale at which the process is operating), and spatial extent (extent of observation). In reference to context, cell-to-cell communication detection or tissue section imputation are representative examples of the scale property.

To conduct the ML of scs, we need to add spatial information to the sc omics data, taking advantage of these spatial properties of dependence, heterogeneity and scale described above. Based on the handling of the spatial information, two different ML approaches can be defined [118]: (i) scs ML methods on which spatial properties are developed in the spatial observation matrix without changing the substance of the learning algorithm, and (ii) scs ML methods on which spatial properties are not handled in the input data but in the the learning algorithm.

## 6.2.1 Spatial information included in the observation matrix

When spatial properties are already represented in the observation matrix, ML techniques do not need to consider the impact of spatial dependency in the learning process. This provides a wide range of useful approximations [22]. Parametric regression techniques are used to model relationships between sc omics and observed spatial data through expressions priorly known or derived from first-principles models (eg: logistic growths in populations dynamics, sigmoidal models for protein abundance, differential equations modelling spatial behaviours, polynomial regressions describing temporarily variables genes, negative binomial distributions for data counting processes). Some representative examples are by Andersson et al. [118], who use a prior negative binomial distribution for cell-type topography inference in low resolution spatial data, and Cang et al. [119], who use parametric regression of an optimal multi-species coupling score in cell-to-cell communication inference.

Non-parametric regression provides more flexible approaches for modeling relationships between observed or latent omics variables and spatial information without an explicit analytical expression. Kernel-based strategies model nested hierarchies in the data considering complex covariate data structures. Gaussian Processes (GPs) are one of the most representative kernel methods given their capacity to work with sparse data, accounting for irregular sampling or incompleteness. GPs kernels can, for example, capture spatial correlated variations of expressions of genes between spatial locations, or cluster gene expression profiles [120], [121]. Other spatial kernels modelling spatial cross-correlations [122] or weighted mean of cell expressions in the neighborhood and azimuthal Gabor filters [123], are used to label spatially structured cell types. Regression techniques can have different flavors depending on design parameters like prior knowledge or likelihood (e.g, Bayesian models are used to increase spatial resolution of images [113]), kernel properties, loss functions constraints or bias-variance tradeoff among others.

In the group of graph-based ML modeling, ensembles of Decision Trees like Random Forest (RF) use observations spatial properties extracted with Gaussian filters and their derivatives (e.g. pixels intensity, edge or texture) to model probabilities for supervised cell segmentation [124], [125]. Deep Learning (DL) approaches, a large family of probabilistic models in which differentiable functions are composed into any kind of directed acyclic graph, provide very useful approximations of scs ML. Stacked feed-forward networks, named Deep Neural Networks (DNNs), are the basic instance of DL model that can follow diverse architectures (e.g. auto-encoders for latent space modeling or recurrent neural networks to model sequential data) depending on the needs of the information to be modeled. As an example, Prasad et al. [113] propose a Deep masked auto-encoder for self-supervised coding of latent representation of gene expression embedded with spatial information. Following a different probabilistic approach, special mention should be made to Generative Deep Learning models used to, in brief, model the joint distribution of omics and spatial data. This fact provides them with a good ability to deal with missing data, to increase spatial resolution and to generate samples with certain desired properties by interpolating between existing data points. Variational Autoencoders (VAEs), which follow the encoder-decoder architecture to learn a joint probabilistic distribution of the input and encoded latent space, are often used in scs applications. Fischer et al. [127] provide an example of VAE in collaboration with other architectures, to infer cell-cell dependencies and cell-intrinsic latent states, to capture ligand-receptor based communications and to estimate the effects of niche composition on gene expression.

In connection to the ML approaches reviewed, several critical aspects are spatial sampling, spatial features, dimensionality reduction, and handling of missing data [117]. Advances in these aspects will result in improvements of the algorithms used. When dealing with multi-omics data these aspects grow challenging given the disparity in feature types and dimension, spatial resolution and spatial context.

### 6.2.2 Handling spatial information in the learning algorithm

In this second set of ML strategies, the algorithms are responsible for considering spatial dependencies in the learning process. Graph-based models are intrinsic representatives of this group. They express functional relationships between omics features and space using directed or undirected graphs with nodes corresponding to samples and edges conforming spatial proximity, while conditional independence between samples is encoded. They are widely used in scs, given their capacity to model soft spatial constraints between samples, often under spatially unsupervised knowledge discovery. Pham et al. [128] present a graph-based clustering with the well-known toolkit stLearn to uncover relationships between transcriptional states of cells across tissues undergoing dynamic changes. In this same category, the undirected graph models HMRF has proven enhanced modelling capacities in spatial domains [86], [129], [54].

In Convolutional Neural Networks (CNNs) and in the auto-encoders' homolog U-Net, the matrix multiplications in DNNs are replaced by convolution operations to capture spatial information. Monjo et al. [130] provide an illustrative example of CNNs applied in spatial gene clustering and expression, allowing for the super-resolution prediction of transcript levels at the inter-spot space and at unmeasured spots, and tissue section imputation. Recently Transformers, based on the multi-head attention mechanism, have been proposed to model spatial dependencies in omics data. Xu et al. [108] presents an interesting example of a transformer-based architecture for the integration of single-cell proteomics with other omics data through an ad-hoc defined multi-modal multi-task attention mechanism with promising results.

It is often common to combine graph-based models and Neural Networks to leverage non-linear modeling potential when complex data relationships are present. Applying convolutions in the spatial domain defined by the graphs allows for a more efficient integration of spatial information from neighbouring nodes. A wide range of computational possibilities can be found in the literature. Graph Neural Networks (GNNs) are used for several scs applications like, e.g., identification of interactions through membrane-bound ligand receptor pairs [119], [54] or data coverage increase [112]. Using convolutions through Graph Convolutional Networks (GCN) improves very high-dimensional data modelling. Hu et al [111] use the previous approach in the integration of gene expression, spatial localization, and histology. Fischer et al. [127] use a Conditional Graph Variational Autoencoder to model the cell's intrinsic latent states conditioned over node expression states on a graph embedding of the niche and the cell type. State of the art examples of combining graph-based models and Transformers are proposed by Zhao et al. [131] for the reconstruction of super-resolved gene expressions coupled with a GNN to recover detailed tissue structures in histology images at sc resolution, or by Xiao et al. [132], who combine convolutional layers, transformer encoders, and graph neural networks to estimate gene expressions from histology images.

### 6.2.3 Hybrid approaches

Some of the best practices proposed for handling spatial properties in scs data combine the simultaneous modeling of spatial dependencies and heterogeneities both in the observation matrix and the learning algorithm. This provides a series of graph or CNN tandems with state of the art DL strategies and attention mechanisms under constant improvement to give answers to challenging spatial modelling objectives like, e.g., comprehensive integration of multi-omics spatial data, microenvironment inference or integration of heterogeneous atlas data. Whether to choose or not DL strategies depends on the complexity and prior knowledge of the problem to be modeled, the amount of data available (scarcity can be partially manageable through the usage of pre-trained models and transfer learning), and the computational resources at disposal.

Regarding the structural property of scale of spatial information, Deep Generative Models provide interesting approximations for tackling the need to increase spatial resolution through sampling learnt probability distributions. Being Generative probabilistic models a very active field of work, it is interesting to point at diffusion models which are starting to be used in the field of omics [133] [134].

Simultaneous learning multi-scale spatio-temporal patterns [117] (approachable with combinations of RNNs, CNNs, Transformers or other NN competitive approximations) goes beyond the scope of analysis but it contextualizes the limits of *context* and *extent* scale properties of spatial information.

# 7 Conclusion

This survey presents the state of the art of omics technologies from various biological domains, namely (epi)genomics, transcriptomics, proteomics and metabolomics, which combine both single-cell resolution and spatial information. We generally refer to this data as single-cell spatial (scs) omics. The survey pays special attention to downstream analysis methods, in particular to approaches for the modelling of spatial information. We discuss scs tools and approaches in the literature, and developments in sister areas that can be leveraged to model scs data.

We end this paper by suggesting some relevant future challenges for the analysis of scs data from a computational perspective.

Temporal measurements (repeated measures) are a cornerstone of medical research, useful to unveil temporal patterns in conditions, or the efficacy of treatments. Even though scs technologies are still under development, and the combination of scs data with temporal information is still in its infancy, we forecast that modelling spatio-temporal patterns is the next frontier for downstream analysis tools.

Another interesting direction to enhance the capabilities of analysis methods is the explicit integration of information about the experimental design into the analysis tool. This can be performed either by extending tools like ASCA or PERMANOVA to incorporate spatial, massive and sparse information, or by extending scs methods to accommodate the information of the experimental design.

Finally, the integration of sc omics with imaging modalities beyond standard microscopy (e.g., cryo-electron tomography) and the use of AI-driven approaches in automated real-time data analysis are interesting computational problems for future research.

# Funding acknowledgments

This work was supported by grant no. PID2023-1523010B-IOO (MuSTARD), funded by the Agencia Estatal de Investigación in Spain, call no. MCIN/AEI/10.13039/501100011033, and by the European Regional Development Fund. Michael Sorochan Armstrong was funded through the MSCA program, project: MAHOD-101106986.

# References

[1] T. Hu, J. Li, H. Zhou, C. Li, E. C. Holmes, and W. Shi, 'Bioinformatics resources for SARS-CoV-2 discovery and surveillance', *Brief. Bioinform.*, vol. 22, no. 2, pp. 631–641, Mar. 2021, doi: 10.1093/bib/bbaa386.

[2] S. Alseekh *et al.*, 'Mass spectrometry-based metabolomics: a guide for annotation, quantification and best reporting practices', *Nat. Methods*, vol. 18, no. 7, pp. 747–756, Jul. 2021, doi: 10.1038/s41592-021-01197-1.

[3] T. Stuart *et al.*, 'Comprehensive Integration of Single-Cell Data', *Cell*, vol. 177, no. 7, pp. 1888-1902.e21, Jun. 2019, doi: 10.1016/j.cell.2019.05.031.

[4]     V. Marx, 'Method of the Year: spatially resolved transcriptomics', *Nat. Methods*, vol. 18, no. 1, pp. 9–14, Jan. 2021, doi: 10.1038/s41592-020-01033-y.

[5]     M. Eisenstein, 'Seven technologies to watch in 2022', *Nature*, vol. 601, no. 7894, pp. 658–661, Jan. 2022, doi: 10.1038/d41586-022-00163-x.

[6]     'Top 10 Emerging Technologies of 2023 report', World Economic Forum. Accessed: Apr. 10, 2024. [Online]. Available: https://www.weforum.org/publications/top-10-emerging-technologies-of-2023/

[7]     S. Qin, D. Miao, X. Zhang, Y. Zhang, and Y. Bai, 'Methods developments of mass spectrometry based single cell metabolomics', *TrAC Trends Anal. Chem.*, vol. 164, p. 117086, Jul. 2023, doi: 10.1016/j.trac.2023.117086.

[8]     S. Ma, Y. Leng, X. Li, Y. Meng, Z. Yin, and W. Hang, 'High spatial resolution mass spectrometry imaging for spatial metabolomics: Advances, challenges, and future perspectives', *TrAC Trends Anal. Chem.*, vol. 159, p. 116902, Feb. 2023, doi: 10.1016/j.trac.2022.116902.

[9]     I. Kleino, P. Frolovaitė, T. Suomi, and L. L. Elo, 'Computational solutions for spatial transcriptomics', *Comput. Struct. Biotechnol. J.*, vol. 20, pp. 4870–4884, 2022, doi: 10.1016/j.csbj.2022.08.043.

[10]    Y. Wu, Y. Cheng, X. Wang, J. Fan, and Q. Gao, 'Spatial omics: Navigating to the golden era of cancer research', *Clin. Transl. Med.*, vol. 12, no. 1, p. e696, Jan. 2022, doi: 10.1002/ctm2.696.

[11]    R. Ahmed *et al.*, 'Spatial mapping of cancer tissues by OMICS technologies', *Biochim. Biophys. Acta BBA - Rev. Cancer*, vol. 1877, no. 1, p. 188663, Jan. 2022, doi: 10.1016/j.bbcan.2021.188663.

[12]    L. Tian, F. Chen, and E. Z. Macosko, 'The expanding vistas of spatial transcriptomics', *Nat. Biotechnol.*, vol. 41, no. 6, pp. 773–782, Jun. 2023, doi: 10.1038/s41587-022-01448-2.

[13]    L. Larsson, J. Frisén, and J. Lundeberg, 'Spatially resolved transcriptomics adds a new dimension to genomics', *Nat. Methods*, vol. 18, no. 1, pp. 15–18, Jan. 2021, doi: 10.1038/s41592-020-01038-7.

[14]    B. He *et al.*, 'Assessing the Impact of Data Preprocessing on Analyzing Next Generation Sequencing Data', *Front. Bioeng. Biotechnol.*, vol. 8, Jul. 2020, doi: 10.3389/fbioe.2020.00817.

[15]    Y. Liu, Z. Shao, and N. Hoffmann, 'Global Attention Mechanism: Retain Information to Enhance Channel-Spatial Interactions'. arXiv, Dec. 10, 2021. Accessed: Apr. 01, 2024. [Online]. Available: http://arxiv.org/abs/2112.05561

[16]    P. Mishra, A. Biancolillo, J. M. Roger, F. Marini, and D. N. Rutledge, 'New data preprocessing trends based on ensemble of multiple preprocessing techniques', *TrAC Trends Anal. Chem.*, vol. 132, p. 116045, Nov. 2020, doi: 10.1016/j.trac.2020.116045.

[17]    M. D. Sorochan Armstrong, J. L. Hinrich, A. P. de la Mata, and J. J. Harynuk, 'PARAFAC2×N: Coupled decomposition of multi-modal data with drift in N modes', *Anal. Chim. Acta*, vol. 1249, p. 340909, Apr. 2023, doi: 10.1016/j.aca.2023.340909.

[18]    W. W. B. Goh and L. Wong, 'Dealing with Confounders in Omics Analysis', *Trends Biotechnol.*, vol. 36, no. 5, pp. 488–498, May 2018, doi: 10.1016/j.tibtech.2018.01.013.

[19]    R. Yamada, D. Okada, J. Wang, T. Basak, and S. Koyama, 'Interpretation of omics data analyses', *J. Hum. Genet.*, vol. 66, no. 1, pp. 93–102, Jan. 2021, doi: 10.1038/s10038-020-

0763-5.

[20] S. Tarazona, A. Arzalluz-Luque, and A. Conesa, 'Undisclosed, unmet and neglected challenges in multi-omics studies', *Nat. Comput. Sci.*, vol. 1, no. 6, pp. 395–402, Jun. 2021, doi: 10.1038/s43588-021-00086-z.

[21] D. Lähnemann *et al.*, 'Eleven grand challenges in single-cell data science', *Genome Biol.*, vol. 21, no. 1, p. 31, Feb. 2020, doi: 10.1186/s13059-020-1926-6.

[22] B. Velten and O. Stegle, 'Principles and challenges of modeling temporal and spatial omics data', *Nat. Methods*, vol. 20, no. 10, pp. 1462–1474, Oct. 2023, doi: 10.1038/s41592-023-01992-y.

[23] B. A. M. Bouwman, N. Crosetto, and M. Bienko, 'The era of 3D and spatial genomics', *Trends Genet.*, vol. 38, no. 10, pp. 1062–1075, Oct. 2022, doi: 10.1016/j.tig.2022.05.010.

[24] J. Cao *et al.*, 'Microfluidics-based single cell analysis: from transcriptomics to spatiotemporal multi-omics', *TrAC Trends Anal. Chem.*, vol. 158, p. 116868, Jan. 2023, doi: 10.1016/j.trac.2022.116868.

[25] Z. Wang, M. Cao, S. M. Lam, and G. Shui, 'Embracing lipidomics at single-cell resolution: Promises and pitfalls', *TrAC Trends Anal. Chem.*, vol. 160, p. 116973, Mar. 2023, doi: 10.1016/j.trac.2023.116973.

[26] L. Heumos *et al.*, 'Best practices for single-cell analysis across modalities', *Nat. Rev. Genet.*, vol. 24, no. 8, pp. 550–572, Aug. 2023, doi: 10.1038/s41576-023-00586-w.

[27] Y. Wang *et al.*, 'Spatial transcriptomics: Technologies, applications and experimental considerations', *Genomics*, vol. 115, no. 5, p. 110671, Sep. 2023, doi: 10.1016/j.ygeno.2023.110671.

[28] K. Vandereyken, A. Sifrim, B. Thienpont, and T. Voet, 'Methods and applications for single-cell and spatial multi-omics', *Nat. Rev. Genet.*, vol. 24, no. 8, pp. 494–515, Aug. 2023, doi: 10.1038/s41576-023-00580-2.

[29] C. G. Williams, H. J. Lee, T. Asatsuma, R. Vento-Tormo, and A. Haque, 'An introduction to spatial transcriptomics for biomedical research', *Genome Med.*, vol. 14, no. 1, p. 68, Jun. 2022, doi: 10.1186/s13073-022-01075-1.

[30] L. Greenstreet *et al.*, 'DNA-GPS: A theoretical framework for optics-free spatial genomics and synthesis of current methods', *Cell Syst.*, vol. 14, no. 10, pp. 844-859.e4, Oct. 2023, doi: 10.1016/j.cels.2023.08.005.

[31] D. J. Burgess, 'Spatial transcriptomics coming of age', *Nat. Rev. Genet.*, vol. 20, no. 6, pp. 317–317, Jun. 2019, doi: 10.1038/s41576-019-0129-z.

[32] L. E. Borm *et al.*, 'Scalable in situ single-cell profiling by electrophoretic capture of mRNA using EEL FISH', *Nat. Biotechnol.*, Sep. 2022, doi: 10.1038/s41587-022-01455-3.

[33] L. Moses and L. Pachter, 'Museum of spatial transcriptomics', *Nat. Methods*, vol. 19, no. 5, pp. 534–546, May 2022, doi: 10.1038/s41592-022-01409-2.

[34] J. Chen, Y. Wang, and J. Ko, 'Single-cell and spatially resolved omics: Advances and limitations', *J. Pharm. Anal.*, vol. 13, no. 8, pp. 833–835, Aug. 2023, doi: 10.1016/j.jpha.2023.07.002.

[35] X. Wang *et al.*, 'Three-dimensional intact-tissue sequencing of single-cell transcriptional states', *Science*, vol. 361, no. 6400, p. eaat5691, Jul. 2018, doi: 10.1126/science.aat5691.

[36] X. Jiang *et al.*, 'A new direction in metabolomics: Analysis of the central nervous system

based on spatially resolved metabolomics', *TrAC Trends Anal. Chem.*, vol. 165, p. 117103, Aug. 2023, doi: 10.1016/j.trac.2023.117103.

[37] M. J. Taylor, J. K. Lukowski, and C. R. Anderton, 'Spatially Resolved Mass Spectrometry at the Single Cell: Recent Innovations in Proteomics and Metabolomics', *J. Am. Soc. Mass Spectrom.*, vol. 32, no. 4, pp. 872–894, Apr. 2021, doi: 10.1021/jasms.0c00439.

[38] L. Xing *et al.*, 'Next Generation of Mass Spectrometry Imaging: from Micrometer to Subcellular Resolution', *Chem. Biomed. Imaging*, vol. 1, no. 8, pp. 670–682, Nov. 2023, doi: 10.1021/cbmi.3c00061.

[39] T. Alexandrov, 'Spatial Metabolomics and Imaging Mass Spectrometry in the Age of Artificial Intelligence', *Annu. Rev. Biomed. Data Sci.*, vol. 3, no. 1, pp. 61–87, Jul. 2020, doi: 10.1146/annurev-biodatasci-011420-031537.

[40] H. Tian, S. Sheraz Née Rabbani, J. C. Vickerman, and N. Winograd, 'Multiomics Imaging Using High-Energy Water Gas Cluster Ion Beam Secondary Ion Mass Spectrometry [$(H_2O)_n$-GCIB-SIMS] of Frozen-Hydrated Cells and Tissue', *Anal. Chem.*, vol. 93, no. 22, pp. 7808–7814, Jun. 2021, doi: 10.1021/acs.analchem.0c05210.

[41] S. Lee, H. M. Vu, J.-H. Lee, H. Lim, and M.-S. Kim, 'Advances in Mass Spectrometry-Based Single Cell Analysis', *Biology*, vol. 12, no. 3, p. 395, Mar. 2023, doi: 10.3390/biology12030395.

[42] B. Wang, K. Yao, and Z. Hu, 'Advances in mass spectrometry-based single-cell metabolite analysis', *TrAC Trends Anal. Chem.*, vol. 163, p. 117075, Jun. 2023, doi: 10.1016/j.trac.2023.117075.

[43] L. Rappez *et al.*, 'SpaceM reveals metabolic states of single cells', *Nat. Methods*, vol. 18, no. 7, pp. 799–805, Jul. 2021, doi: 10.1038/s41592-021-01198-0.

[44] K. Karrobi *et al.*, 'Fluorescence Lifetime Imaging Microscopy (FLIM) reveals spatial-metabolic changes in 3D breast cancer spheroids', *Sci. Rep.*, vol. 13, no. 1, p. 3624, Mar. 2023, doi: 10.1038/s41598-023-30403-7.

[45] L. M. Dowling *et al.*, 'Fourier Transform Infrared microspectroscopy identifies single cancer cells in blood. A feasibility study towards liquid biopsy', *PLOS ONE*, vol. 18, no. 8, p. e0289824, Aug. 2023, doi: 10.1371/journal.pone.0289824.

[46] L. Yue *et al.*, 'A guidebook of spatial transcriptomic technologies, data resources and analysis approaches', *Comput. Struct. Biotechnol. J.*, vol. 21, pp. 940–955, 2023, doi: 10.1016/j.csbj.2023.01.016.

[47] P. Ràfols *et al.*, 'Signal preprocessing, multivariate analysis and software tools for MA(LDI)-TOF mass spectrometry imaging for biological applications', *Mass Spectrom. Rev.*, vol. 37, no. 3, pp. 281–306, 2018, doi: 10.1002/mas.21527.

[48] G. Baquer, L. Sementé, T. Mahamdi, X. Correig, P. Ràfols, and M. García-Altares, 'What are we imaging? Software tools and experimental strategies for annotation and identification of small molecules in mass spectrometry imaging', *Mass Spectrom. Rev.*, vol. 42, no. 5, pp. 1927–1964, 2023, doi: 10.1002/mas.21794.

[49] W. D. Cameron, A. M. Bennett, C. V. Bui, H. H. Chang, and J. V. Rocheleau, 'Leveraging multimodal microscopy to optimize deep learning models for cell segmentation', *APL Bioeng.*, vol. 5, no. 1, p. 016101, Jan. 2021, doi: 10.1063/5.0027993.

[50] J. Park *et al.*, 'Cell segmentation-free inference of cell types from in situ transcriptomics data', *Nat. Commun.*, vol. 12, no. 1, p. 3545, Jun. 2021, doi: 10.1038/s41467-021-23807-4.

[51] Y. Hao *et al.*, 'Integrated analysis of multimodal single-cell data', *Cell*, vol. 184, no. 13, pp. 3573-3587.e29, Jun. 2021, doi: 10.1016/j.cell.2021.04.048.

[52] G. Palla *et al.*, 'Squidpy: a scalable framework for spatial omics analysis', *Nat. Methods*, vol. 19, no. 2, pp. 171–178, Feb. 2022, doi: 10.1038/s41592-021-01358-2.

[53] F. A. Wolf, P. Angerer, and F. J. Theis, 'SCANPY: Large-scale single-cell gene expression data analysis', *Genome Biol.*, vol. 19, no. 1, 2018, doi: 10.1186/s13059-017-1382-0.

[54] R. Dries *et al.*, 'Giotto: a toolbox for integrative analysis and visualization of spatial expression data', *Genome Biol.*, vol. 22, no. 1, p. 78, Dec. 2021, doi: 10.1186/s13059-021-02286-2.

[55] T. Hu *et al.*, 'Single-cell spatial metabolomics with cell-type specific protein profiling for tissue systems biology', *Nat. Commun.*, vol. 14, no. 1, p. 8260, Dec. 2023, doi: 10.1038/s41467-023-43917-5.

[56] K. Clifton, M. Anant, G. Aihara, L. Atta, O. K. Aimiuwu, J. M. Kebschull, M. I. Miller, D. Tward, and J. Fan, "STalign: Alignment of Spatial Transcriptomics Data Using Diffeomorphic Metric Mapping," *Nat. Commun.*, vol. 14, no. 1, Art. no. 8123, 2023, doi: 10.1038/s41467-023-43915-7.

[57] H. Li, Y. Lin, W. He, W. Han, X. Xu, C. Xu, E. Gao, H. Zhao, and X. Gao, "SANTO: A Coarse-to-Fine Alignment and Stitching Method for Spatial Omics," *Nature Communications*, vol. 15, no. 1, p. 6048, 2024, doi: 10.1038/s41467-024-50308-x.

[58] K. M. Stouffer, A. Trouvé, L. Younes, M. Kunst, L. Ng, H. Zeng, M. Anant, *et al.*, "Cross-Modality Mapping Using Image Varifolds to Align Tissue-Scale Atlases to Molecular-Scale Measures with Application to 2D Brain Sections," *Nature Communications*, vol. 15, no. 1, p. 3530, 2024, doi: 10.1038/s41467-024-47883-4.

[59] Z. Tang, S. Luo, H. Zeng, J. Huang, X. Sui, M. Wu, and X. Wang, "Search and Match across Spatial Omics Samples at Single-Cell Resolution," *Nature Methods*, vol. 21, no. 10, pp. 1818-1829, 2024, doi: 10.1038/s41592-024-02410-7.

[60] M. Koutrouli, E. Karatzas, D. Paez-Espino, and G. A. Pavlopoulos, 'A Guide to Conquer the Biological Network Era Using Graph Theory', *Front. Bioeng. Biotechnol.*, vol. 8, p. 34, Jan. 2020, doi: 10.3389/fbioe.2020.00034.

[61] L. McInnes, J. Healy, N. Saul, and L. Großberger, 'UMAP: Uniform Manifold Approximation and Projection', *J. Open Source Softw.*, vol. 3, no. 29, p. 861, Sep. 2018, doi: 10.21105/joss.00861.

[62] A. K. Smilde, T. Næs, and K. H. Liland, *Multiblock Data Fusion in Statistics and Machine Learning: Applications in the Natural and Life Sciences*. John Wiley & Sons, 2022.

[63] J. D. Welch, V. Kozareva, A. Ferreira, C. Vanderburg, C. Martin, and E. Z. Macosko, "Single-Cell Multi-Omic Integration Compares and Contrasts Features of Brain Cell Identity," *Cell*, vol. 177, no. 7, pp. 1873–1887.e17, 2019, doi: 10.1016/j.cell.2019.05.006.

[64] K. R. Moon, D. Van Dijk, Z. Wang, S. Gigante, D. B. Burkhardt, W. S. Chen, K. Yim, *et al.*, "Visualizing structure and transitions in high-dimensional biological data," *Nature Biotechnology*, vol. 37, no. 12, pp. 1482–1492, 2019.

[65] J. Bergenstråhle, L. Larsson, and J. Lundeberg, "Seamless Integration of Image and Molecular Analysis for Spatial Transcriptomics Workflows," *BMC Genomics*, vol. 21, no. 1, p. 482, 2020, doi: 10.1186/s12864-020-06832-3.

[66] I. Jolliffe, 'Principal Component Analysis for Special Types of Data', in *Principal*

*Component Analysis*, in Springer Series in Statistics. , New York: Springer-Verlag, 2002, pp. 338–372. doi: 10.1007/0-387-22440-8_13.

[67] D. R. Hardoon, S. Szedmak, and J. Shawe-Taylor, 'Canonical correlation analysis: An overview with application to learning methods', *Neural Comput.*, vol. 16, no. 12, pp. 2639–2664, 2004.

[68] S. Wold, M. Sjöström, and L. Eriksson, 'PLS-regression: a basic tool of chemometrics', *Chemom. Intell. Lab. Syst.*, vol. 58, no. 2, pp. 109–130, 2001.

[69] M. Barker and W. Rayens, 'Partial least squares for discrimination', *J. Chemom.*, vol. 17, no. 3, pp. 166–173, Mar. 2003, doi: 10.1002/cem.785.

[70] H. Zou, T. Hastie, and R. Tibshirani, 'Sparse Principal Component Analysis', *J. Comput. Graph. Stat.*, vol. 15, no. 2, pp. 265–286, Jun. 2006, doi: 10.1198/106186006X113430.

[71] K.-A. Lê Cao, D. Rossouw, C. Robert-Granié, and P. Besse, 'A Sparse PLS for Variable Selection when Integrating Omics Data', *Stat. Appl. Genet. Mol. Biol.*, vol. 7, no. 1, Jan. 2008, doi: 10.2202/1544-6115.1390.

[72] S. Park, E. Ceulemans, and K. Van Deun, 'Sparse common and distinctive covariates regression', *J. Chemom.*, vol. 35, no. 2, Feb. 2021, doi: 10.1002/cem.3270.

[73] D. Lee and H. S. Seung, 'Algorithms for non-negative matrix factorization', *Adv. Neural Inf. Process. Syst.*, vol. 13, 2000, Accessed: Apr. 12, 2024. [Online]. Available: https://proceedings.neurips.cc/paper_files/paper/2000/hash/f9d1152547c0bde01830b7e8bd60024c-Abstract.html

[74] I. Borg and P. J. F. Groenen, *Modern Multidimensional Scaling: Theory and Applications*. Springer Science & Business Media, 2005.

[75] L. van der Maaten and G. Hinton, 'Visualizing Data using t-SNE', *J. Mach. Learn. Res.*, vol. 9, no. 86, pp. 2579–2605, 2008.

[76] E. Szymańska, E. Saccenti, A. K. Smilde, and J. A. Westerhuis, 'Double-check: validation of diagnostic statistics for PLS-DA models in metabolomics studies', *Metabolomics*, vol. 8, no. 1, pp. 3–16, Jun. 2012, doi: 10.1007/s11306-011-0330-3.

[77] M. A. Myers, S. Zaccaria, and B. J. Raphael, 'Identifying tumor clones in sparse single-cell mutation data', *Bioinformatics*, vol. 36, no. Supplement_1, pp. i186–i193, Jul. 2020, doi: 10.1093/bioinformatics/btaa449.

[78] W. S. Torgerson, 'Multidimensional scaling: I. Theory and method', *Psychometrika*, vol. 17, no. 4, pp. 401–419, Dec. 1952, doi: 10.1007/BF02288916.

[79] D. Kobak and P. Berens, 'The art of using t-SNE for single-cell transcriptomics', *Nat. Commun.*, vol. 10, no. 1, p. 5416, Nov. 2019, doi: 10.1038/s41467-019-13056-x.

[80] V. H. Do and S. Canzar, 'A generalization of t-SNE and UMAP to single-cell multimodal omics', *Genome Biol.*, vol. 22, no. 1, p. 130, Dec. 2021, doi: 10.1186/s13059-021-02356-5.

[81] T. Smets *et al.*, 'Evaluation of Distance Metrics and Spatial Autocorrelation in Uniform Manifold Approximation and Projection Applied to Mass Spectrometry Imaging Data', *Anal. Chem.*, vol. 91, no. 9, pp. 5706–5714, May 2019, doi: 10.1021/acs.analchem.8b05827.

[82] A. Sarycheva *et al.*, 'Structure-Preserving and Perceptually Consistent Approach for Visualization of Mass Spectrometry Imaging Datasets', *Anal. Chem.*, vol. 93, no. 3, pp. 1677–1685, Jan. 2021, doi: 10.1021/acs.analchem.0c04256.

[83] H. Hu, R. Yin, H. M. Brown, and J. Laskin, 'Spatial Segmentation of Mass Spectrometry

Imaging Data by Combining Multivariate Clustering and Univariate Thresholding', *Anal. Chem.*, vol. 93, no. 7, pp. 3477–3485, Feb. 2021, doi: 10.1021/acs.analchem.0c04798.

[84] L. Shang and X. Zhou, 'Spatially aware dimension reduction for spatial transcriptomics', *Nat. Commun.*, vol. 13, no. 1, p. 7203, 2022.

[85] F. W. Townes and B. E. Engelhardt, 'Nonnegative spatial factorization applied to spatial genomics', *Nat. Methods*, vol. 20, no. 2, pp. 229–238, Feb. 2023, doi: 10.1038/s41592-022-01687-w.

[86] B. Chidester, T. Zhou, S. Alam, and J. Ma, 'SpiceMix enables integrative single-cell spatial modeling of cell identity', *Nat. Genet.*, vol. 55, no. 1, pp. 78–88, Jan. 2023, doi: 10.1038/s41588-022-01256-z.

[87] L. Moses *et al.*, 'Voyager: exploratory single-cell genomics data analysis with geospatial statistics', *bioRxiv*, p. 2023.07.20.549945, Aug. 2023, doi: 10.1101/2023.07.20.549945.

[88] J. M. Prats-Montalbán, A. De Juan, and A. Ferrer, 'Multivariate image analysis: A review with applications', *Chemom. Intell. Lab. Syst.*, vol. 107, no. 1, pp. 1–23, May 2011, doi: 10.1016/j.chemolab.2011.03.002.

[89] M. H. Bharati and J. F. MacGregor, 'Texture analysis of images using principal component analysis', in *Process Imaging for Automatic Control*, SPIE, Feb. 2001, pp. 27–37. doi: 10.1117/12.417179.

[90] A. de Juan and R. Tauler, 'Multivariate Curve Resolution: 50 years addressing the mixture analysis problem – A review', *Anal. Chim. Acta*, vol. 1145, pp. 59–78, Feb. 2021, doi: 10.1016/j.aca.2020.10.051.

[91] C. Ruckebusch and L. Blanchet, 'Multivariate curve resolution: A review of advanced and tailored applications and challenges', *Anal. Chim. Acta*, vol. 765, pp. 28–36, Feb. 2013, doi: 10.1016/j.aca.2012.12.028.

[92] J. Jaumot and R. Tauler, 'Potential use of multivariate curve resolution for the analysis of mass spectrometry images', *The Analyst*, vol. 140, no. 3, pp. 837–846, 2015, doi: 10.1039/C4AN00801D.

[93] S. Hugelier, O. Devos, and C. Ruckebusch, 'On the implementation of spatial constraints in multivariate curve resolution alternating least squares for hyperspectral image analysis', *J. Chemom.*, vol. 29, no. 10, pp. 557–561, 2015, doi: 10.1002/cem.2742.

[94] P. Firmani, S. Hugelier, F. Marini, and C. Ruckebusch, 'MCR-ALS of hyperspectral images with spatio-spectral fuzzy clustering constraint', *Chemom. Intell. Lab. Syst.*, vol. 179, pp. 85–91, Aug. 2018, doi: 10.1016/j.chemolab.2018.06.007.

[95] R. Vitale, S. Hugelier, D. Cevoli, and C. Ruckebusch, 'A spatial constraint to model and extract texture components in Multivariate Curve Resolution of near-infrared hyperspectral images', *Anal. Chim. Acta*, vol. 1095, pp. 30–37, Jan. 2020, doi: 10.1016/j.aca.2019.10.028.

[96] M. Li Vigni, J. M. Prats-Montalban, A. Ferrer, and M. Cocchi, 'Coupling 2D-wavelet decomposition and multivariate image analysis (2D WT-MIA)', *J. Chemom.*, vol. 32, no. 1, p. e2970, Jan. 2018, doi: 10.1002/cem.2970.

[97] S. M. Zandavi *et al.*, 'Disentangling single-cell omics representation with a power spectral density-based feature extraction', *Nucleic Acids Res.*, vol. 50, no. 10, pp. 5482–5492, Jun. 2022, doi: 10.1093/nar/gkac436.

[98] Y.-L. Gao, Q. Qiao, J. Wang, S.-S. Yuan, and J.-X. Liu, 'BioSTD: A New Tensor Multi-View

Framework via Combining Tensor Decomposition and Strong Complementarity Constraint for Analyzing Cancer Omics Data', *IEEE J. Biomed. Health Inform.*, vol. 27, no. 10, pp. 5187–5198, Oct. 2023, doi: 10.1109/JBHI.2023.3299274.

[99] A. A. Stepanchuk and P. K. Stys, 'Spectral Fluorescence Pathology of Protein Misfolding Disorders', *ACS Chem. Neurosci.*, vol. 15, no. 5, pp. 898–908, Mar. 2024, doi: 10.1021/acschemneuro.3c00798.

[100] M. Kircheis and D. Potts, 'Direct inversion of the nonequispaced fast Fourier transform', *Linear Algebra Its Appl.*, vol. 575, pp. 106–140, Aug. 2019, doi: 10.1016/j.laa.2019.03.028.

[101] M. S. Armstrong, J. C. Pérez-Girón, J. Camacho, and R. Zamora, 'A direct solution to the interpolative inverse non-uniform fast Fourier transform problem for spectral analyses of non-equidistant time-series data'. arXiv, Feb. 27, 2024. Accessed: Apr. 12, 2024. [Online]. Available: http://arxiv.org/abs/2310.15310

[102] D. Montgomery, *Design and Analysis of Experiments*. Wiley, 2020.

[103] M. J. Anderson, "Permutational multivariate analysis of variance (PERMANOVA)," *Wiley StatsRef: Statistics Reference Online*, pp. 1–15, 2014.

[104] A. K. Smilde, J. J. Jansen, H. C. Hoefsloot, R.-J. A. Lamers, J. Van Der Greef, and M. E. Timmerman, "Anova-simultaneous component analysis (asca): A new tool for analyzing designed metabolomics data," *Bioinformatics*, vol. 21, no. 13, pp. 3043–3048, 2005.

[105] B. J. Kelly, R. Gross, K. Bittinger, S. Sherrill-Mix, J. D. Lewis, R. G. Collman, F. D. Bushman, and H. Li, "Power and sample-size estimation for microbiome studies using pairwise distances and PERMANOVA," *Bioinformatics*, vol. 31, no. 15, pp. 2461–2468, Aug. 2015, doi: 10.1093/bioinformatics/btv183.

[106] J. Camacho and M. S. Armstrong, "Population Power Curves in ASCA With Permutation Testing," *Journal of Chemometrics*, 2024, doi: e3596.

[107] H. Chen, D. Li, and Z. Bar-Joseph, 'SCS: cell segmentation for high-resolution spatial transcriptomics', *Nat. Methods*, vol. 20, no. 8, pp. 1237–1243, Aug. 2023, doi: 10.1038/s41592-023-01939-3.

[108] J. Xu, D. Huang, and X. Zhang, 'scmFormer Integrates Large-Scale Single-Cell Proteomics and Transcriptomics Data by Multi-Task Transformer', *Adv. Sci.*, p. 2307835, Mar. 2024, doi: 10.1002/advs.202307835.

[109] T. Biancalani *et al.*, 'Deep learning and alignment of spatially resolved single-cell transcriptomes with Tangram', *Nat. Methods*, vol. 18, no. 11, pp. 1352–1362, Nov. 2021, doi: 10.1038/s41592-021-01264-7.

[110] Y. Lin, T.-Y. Wu, S. Wan, J. Y. H. Yang, W. H. Wong, and Y. X. R. Wang, 'scJoint integrates atlas-scale single-cell RNA-seq and ATAC-seq data with transfer learning', *Nat. Biotechnol.*, vol. 40, no. 5, pp. 703–710, May 2022, doi: 10.1038/s41587-021-01161-6.

[111] J. Hu *et al.*, 'SpaGCN: Integrating gene expression, spatial location and histology to identify spatial domains and spatially variable genes by graph convolutional network', *Nat. Methods*, vol. 18, no. 11, pp. 1342–1351, Nov. 2021, doi: 10.1038/s41592-021-01255-8.

[112] D. Pham *et al.*, 'stLearn: integrating spatial location, tissue morphology and gene expression to find cell types, cell-cell interactions and spatial trajectories within undissociated tissues'. May 31, 2020. doi: 10.1101/2020.05.31.125658.

[113] E. Zhao *et al.*, 'Spatial transcriptomics at subspot resolution with BayesSpace', *Nat. Biotechnol.*, vol. 39, no. 11, pp. 1375–1384, Nov. 2021, doi: 10.1038/s41587-021-00935-2.

[114] Y. Long *et al.*, 'Spatially informed clustering, integration, and deconvolution of spatial transcriptomics with GraphST', *Nat. Commun.*, vol. 14, no. 1, p. 1155, Mar. 2023, doi: 10.1038/s41467-023-36796-3.

[115] N. Erfanian *et al.*, 'Deep learning applications in single-cell genomics and transcriptomics data analysis', *Biomed. Pharmacother.*, vol. 165, p. 115077, Sep. 2023, doi: 10.1016/j.biopha.2023.115077.

[116] L. Atta and J. Fan, 'Computational challenges and opportunities in spatially resolved transcriptomic data analysis', *Nat. Commun.*, vol. 12, no. 1, p. 5283, Sep. 2021, doi: 10.1038/s41467-021-25557-9.

[117] B. Nikparvar and J.-C. Thill, 'Machine Learning of Spatial Data', ISPRS Int. J. Geo-Inf., vol. 10, no. 9, Art. no. 9, Sep. 2021, doi: 10.3390/ijgi10090600.

[118] A. Andersson et al., 'Single-cell and spatial transcriptomics enables probabilistic inference of cell type topography', Commun. Biol., vol. 3, no. 1, p. 565, Oct. 2020, doi: 10.1038/s42003-020-01247-y.

[119] Z. Cang et al., 'Screening cell–cell communication in spatial transcriptomics via collective optimal transport', Nat. Methods, vol. 20, no. 2, pp. 218–228, Feb. 2023, doi: 10.1038/s41592-022-01728-4.

[120] V. Svensson, S. A. Teichmann, and O. Stegle, 'SpatialDE: identification of spatially variable genes', Nat. Methods, vol. 15, no. 5, pp. 343–346, May 2018, doi: 10.1038/nmeth.4636.

[121] L. M. Weber, A. Saha, A. Datta, K. D. Hansen, and S. C. Hicks, 'nnSVG for the scalable identification of spatially variable genes using nearest-neighbor Gaussian processes', Nat. Commun., vol. 14, no. 1, p. 4059, Jul. 2023, doi: 10.1038/s41467-023-39748-z.

[122] B. F. Miller, D. Bambah-Mukku, C. Dulac, X. Zhuang, and J. Fan, 'Characterizing spatial gene expression heterogeneity in spatially resolved single-cell transcriptomic data with nonuniform cellular densities', Genome Res., vol. 31, no. 10, pp. 1843–1855, Jan. 2021, doi: 10.1101/gr.271288.120.

[123] V. Singhal et al., 'BANKSY unifies cell typing and tissue domain segmentation for scalable spatial omics data analysis', Nat. Genet., vol. 56, no. 3, pp. 431–441, Mar. 2024, doi: 10.1038/s41588-024-01664-3.

[124] S. Berg, D. Kutra, T. Kroeger, *et al.*, "Ilastik: Interactive Machine Learning for (Bio)Image Analysis," *Nature Methods*, vol. 16, pp. 1226–1232, 2019, doi: 10.1038/s41592-019-0582-9.

[125] S. Vickovic, B. Lötstedt, J. Klughammer, *et al.*, "SM-Omics is an automated platform for high-throughput spatial multi-omics," *Nature Communications*, vol. 13, p. 795, 2022, doi: 10.1038/s41467-022-28445-y.

[126] M. Prasad, G. Postma, P. Franceschi, L. M. C. Buydens, and J. J. Jansen, 'Evaluation and comparison of unsupervised methods for the extraction of spatial patterns from mass spectrometry imaging data (MSI)', Sci. Rep., vol. 12, no. 1, p. 15687, Sep. 2022, doi: 10.1038/s41598-022-19365-4.

[127] D. S. Fischer, A. C. Schaar, and F. J. Theis, 'Modeling intercellular communication in

tissues using spatial graphs of cells', Nat. Biotechnol., vol. 41, no. 3, pp. 332–336, Mar. 2023, doi: 10.1038/s41587-022-01467-z.

[128] D. Pham et al., 'Robust mapping of spatiotemporal trajectories and cell–cell interactions in healthy and diseased tissues', Nat. Commun., vol. 14, no. 1, p. 7739, Nov. 2023, doi: 10.1038/s41467-023-43120-6.

[129] Q. Zhu, S. Shah, R. Dries, L. Cai, and G.-C. Yuan, 'Identification of spatially associated subpopulations by combining scRNAseq and sequential fluorescence in situ hybridization data', Nat. Biotechnol., vol. 36, no. 12, pp. 1183–1190, Dec. 2018, doi: 10.1038/nbt.4260.

[130] T. Monjo, M. Koido, S. Nagasawa, Y. Suzuki, and Y. Kamatani, 'Efficient prediction of a spatial transcriptomics profile better characterizes breast cancer tissue sections without costly experimentation', Sci. Rep., vol. 12, no. 1, p. 4133, Mar. 2022, doi: 10.1038/s41598-022-07685-4.

[131] C. Zhao et al., 'Innovative super-resolution in spatial transcriptomics: a transformer model exploiting histology images and spatial gene expression', Brief. Bioinform., vol. 25, no. 2, p. bbae052, Mar. 2024, doi: 10.1093/bib/bbae052.

[132] X. Xiao, Y. Kong, R. Li, Z. Wang, and H. Lu, 'Transformer with convolution and graph-node co-embedding: An accurate and interpretable vision backbone for predicting gene expressions from local histopathological image', Med. Image Anal., vol. 91, p. 103040, Jan. 2024, doi: 10.1016/j.media.2023.103040.

[133] Z. Guo et al., 'Diffusion Models in Bioinformatics: A New Wave of Deep Learning Revolution in Action', 2023, doi: 10.48550/ARXIV.2302.10907.

[134] M. Sadria and A. Layton, 'The Power of Two: integrating deep diffusion models and variational autoencoders for single-cell transcriptomics analysis'. Apr. 16, 2023. doi: 10.1101/2023.04.13.536789.